# A Mathematical Approach to the Sleep - Waking Cycle

Michael Rvachov (M.A.Rvachev)

## *Abstract*

Three abilities – the ability to recognize sounds, the ability to visually recognize movement and the ability to keep an upright standing position – can function only with using precise measurements of the short time intervals. Other features that these abilities share are that all three are crucial for the survival and – despite this – they are turned off simultaneously during sleep. Instead, presumably, if turning each of them off periodically for a resetting is unavoidable, then doing this one at a time would be the evolutionary choice, if that were possible. This hints that all three abilities share the same time-interval measuring mechanism and this mechanism is what cannot work without a periodical resetting. Another indication that such a mechanism is shared across the whole nervous system is the ubiquity of Pavlovian conditioning. A high level theory is proposed about how such a measuring mechanism must be implemented, in the simplest way, from the point of view of the data management, and how duration can be measured by resonance. If the theory is true it could explain the purpose of the observed changes of the firing frequency of cortical neurons, why periodic sleeping is an unavoidable necessity and what might be the cause of the different stages of sleep and of yawning. The theory provides simpler and less speculative explanations of a number of the experimental results than interpretations by the existing theories (the two-stage model of memory consolidation etc.).

## *Introduction*

The research about the purpose of sleep is countless. We'll assume that the following quote from H. Craig Heller summarizes the state of the research: "Many hypotheses have been put forward for the functions of sleep, but none have been either proved or falsified… Two major lines of thinking about sleep function are that they are involved in the restoration of brain energy metabolism or that they are involved in processes of synapse and circuit maintenance including the consolidation of memories. These functions are not mutually exclusive, and evidence in support of them is discussed. A critical question that must be asked when considering any putative function of sleep is – Why must the brain be taken off-line to accomplish that function?" [1].

We will try to find the answer to this critical question from consideration of how some wakefulness activities can function in principle. This will lead us to the role of the inner time in the animal functioning. Again, there is much discussion about the inner time, in the temporal range from few milliseconds to lunar rhythms and research range from studies of protein oscillation in vitro [42] to philosophical pondering about the time perception and models of temporal consciousness [43]. It seems that the closest to our research reference is described by the following quote from Marc Wittmann at al.: "There is yet no conclusive answer as to what kind of time keeping mechanism is implemented in the brain. Cognitive models, however, which assume that an internal clock with a pacemaker produces subjective time units have been influential in interpreting human time perception and animal timing behavior. In these models, duration is defined as a function of accumulated units over a time span." [44].

These models might be applicable for studying only one or very few time intervals at a time. However an animal is "bombarded" by thousands of incoming events (stimuli) per second. It is not known upfront which one of these stimuli begins a temporarily important events sequence, which means that for each incoming stimulus its own time unit accumulator has to run, at least for few seconds. This would mean that a very large number of the standalone accumulators will have to run simultaneously, which would be very costly.

Instead, we suggest that there is only one – or very few, specific to the sensory inputs (or other specific activities) and synchronized to each other – accumulator that starts its run at the beginning of the wake cycle. The current value of the accumulator is associated with each event as the event time-tag, and there is a mechanism, a comparator[1], that allows comparing of the time-tags to the current time, when they are close enough. This, by itself, it less costly than using the standalone accumulators, at least because comparing the results of the standalone accumulators still requires the comparators. Yet, we speculate further and suggest that the comparator might be resonance based and sort, in one shot, by the forced oscillation amplitude, all the recent events by their time distance from now, emphasizing the recent events and attenuating the old ones. We suggest how such a time bearing mechanism could have been implemented and what could be its evolutionary roots.

Thoroughly going through all the research we could reach online, we have not found anything that could be considered a predecessor of our approach. Yet, as it is shown below, some of the published experimental results suggest that our theory might be correct, though the experiments were not designed specifically to verify the theory.

To narrow the scope of the discussion we'll talk only about humans, though most of what is discussed is applicable to many other animal species.

## *Time Intervals Pinned activities and Snippet Inner Time*

We exist in time, all our activities occur in time and in this sense are time dependent. However for some activities the time dependence is quite specific. There are several activities that cannot function without an inner clock that measures[2] accurately at least short – few milliseconds long – intervals of time. Let us call them "time interval pinned (TIP) activities", to tell them apart from other time dependent activities. The following are the most apparent examples of the TIP activities.

We are capable of assessing velocities of moving objects, predicting their future positions (within some time snippets) and our own movements and efforts which are required to intercept or avoid moving objects. To estimate the velocities we need the ability to remember, compare and differentiate (at least) two positions of the objects. We also need the ability to assess the time intervals between the positions.

---

[1] Maybe it is better to call it "a differentiator" as it finds the difference between now, the present moment, and the time-tag.

[2] Measuring here means comparing (shorter/longer or equal) of the time intervals. This, in turn, requires an ability to remember and reproduce (recall, e.g. for comparing) at least two intervals at a time.

In most cases the positions of the objects are registered based on visual clues. The position differentiation starts with the excitation of different light sensitive cells and with changes in the eye orientation. We already have some detailed understanding of how the visual input is processed in the brain visual cortex. At least, the visual recognition systems that emulate the steps observed in the brains of humans and monkeys work with reasonable success [2]. However, how the time intervals between the positions are registered and measured is not known.

A similar situation occurs with sound recognition. We are capable of recognizing sound sequences, like bird songs, the roar of tigers or human speech. Different incoming sound frequencies excite different hair cells and thus are differentiated. However for the recognition of a sound sequence the duration of each frequency matters. Changing the duration of each frequency can distort a word beyond recognition, even if the order of the frequencies is kept the same [e.g. 3]. It is not known how we register and measure the duration of sound frequencies.

What we call "muscle memory" is also not possible without accurate knowledge of the real time intervals. These muscles move objects of a certain mass (including the body parts). According to Newton's second law the velocity increment is proportional to the force and its action time. To achieve the required motor sequence each muscle must produce a specific force over a specific time interval. The diversity of the possible motor sequences that can be produced by the same muscles means that the sequence is not hardcoded, since the current state does not uniquely define the next state (as in the case of the heartbeat or movements of the digestive system). Somehow the force-time sequences are acquired, memorized and reproduced. The reproduced muscle memory sequences can also be continuously corrected, based on visual, kinesthetic or other cues.

One example of how the muscle memory is used is in our ability to maintain the upright standing position. It is worthwhile considering this example separately in the context of the theories that suggest that sleep is needed for the specific processing of information. While we sleep presumably it is necessary that our sight and hearing be blocked in order for this important data processing to take place. However, why is it necessary for us to sleep lying down? Why do we not sleep in the standing position, since for safety reasons we would be better able to respond to danger, ready to fight or flee? What are the changes in our bodies that are induced by standing that can only be restored by sleeping? Our muscles get tired, but to rest them we do not need to sleep, nor do we need to lie down and certainly not for 8 hours; a few minutes of sitting would help. Do we, just by standing, take in so much information that it requires some special processing? Does the information learned by standing justify 8 hour blackouts for processing it?

All the abilities listed above are TIP activities; they cannot function without a mechanism for measuring time intervals, at least those of short duration.

How are the times at which the events occur recorded and differentiated? There must be parts of the nervous system that register these times or at least the time intervals between events which are close in time. The diverse neuro-oscillators that have been experimentally observed may be involved, though an oscillation counter would be needed to measure random time intervals using a constant frequency oscillator serving as a real biological clock. It seems that so far there are no suggestions as to how such a counter could be implemented biologically.

Setting aside the unknown implementation of the biological clock, let us discuss how it could work from the point of view of the temporal data management.

Let us call an "event" (stimulus) any change that occurs to us (e.g. a change in the visual input, hearing input, muscle strain etc.). For the TIP activity related events (TIP events) A and B to be pinned in time, within some time snippet (~500ms? [4]), the time interval between them must be measured internally with a precision that is adequate for the activity. For this there must exist an internal mechanism (or maybe many mechanisms) for measuring time intervals. Each event must have a handle to which the mechanism can be attached. Let us call the handle the "event time-tag".

For these time-tags there are two possibilities, the "constant" (or "absolute") time-tags that do not change over time and the "variable" (or "relative") tags that do change.

Let us consider 3 obvious examples.

1. Suppose that the events are organized as a queue (of some limited length): when each new event arrives it takes place at the end of the queue while the first event is deleted from the queue. Suppose that the time-tag of the event is its number in the queue; it will be a variable time-tag, because the event's number in the queue will change each time that a new event arrives. The time interval measuring mechanism would calculate the number of events in the queue between A and B. For the queue based mechanism to produce accurate results the events must arrive at a steady pace (constant frequency of the events).
2. Suppose that at the moment when event A arrives an "internal stopwatch" is started. Then the stopwatch itself is both the variable time-tag and the time interval measuring mechanism. The stopwatch reading when event B arrives is the time interval between A and B.[3]
3. Suppose that when the event arrives the current value of some "inner time" is set as the event's constant time-tag. The time interval measuring mechanism should be able to convert the inner times at the moments of arrival of A and B into an assessment of the real time interval between A and B.

Which, if any, of these 3 mechanisms might be actually used?

Within the time snippet needed for the TIP activity (~500 ms), there may be a very large number of TIP events. If variable time-tags are used, as hypothesized in the examples 1–2 or any other possible implementation, these time-tags must be modified in sync with real time to achieve accurate time interval readings. The cost of this is most likely many times higher than the cost of using a constant time-tag time interval measuring mechanism. In particularly, the cost will be high for the stopwatch-type mechanisms, because for each event that is still temporarily relevant there must be its own stopwatch attached to the event.

---

[3] Any mechanism that starts accumulation of the passed time at the moment of the event arrival falls into the stopwatch category. Examples of the abstract models of the "internal stopwatch" used in cognitive psychology are the "internal clock model" or the water clock ("klepsydra") used in time reproduction models [41].

Then let us hypothesize that the constant time-tag time interval measuring mechanism, which sets the value of some "inner time" (at the moment of the event arrival) as the time-tag, is used, and analyze the consequences of this assumption. First of all, let us explore what is the minimum set of properties the unknown "inner time" process should have so that it could serve as a source of the time tags.

Inside us, within the time snippet of the TIP activity, the intervals' "internal images" must be processed with the same proportionality as occurs in the real time intervals. For this the internal process that serves as the constant time-tag must have within the snippet a scalar property that changes proportionally to the real time, Fig.1a. Then the increments of the process will be "convertible" back into the increments of real time and thus usable for the internal time increment assessment. To measure the time duration up to 500 ms the process increments must be proportional to the real time increments at least within the 500 ms time snippets. But since the measurements are happening at *any* moment, these 500 ms snippets will be overlapping each other and covering the whole period when the measurements are occurring. Therefore, for the entire duration when the TIP activities are occurring, the "inner time" process must be (almost) proportional to the real time, Fig.1b.

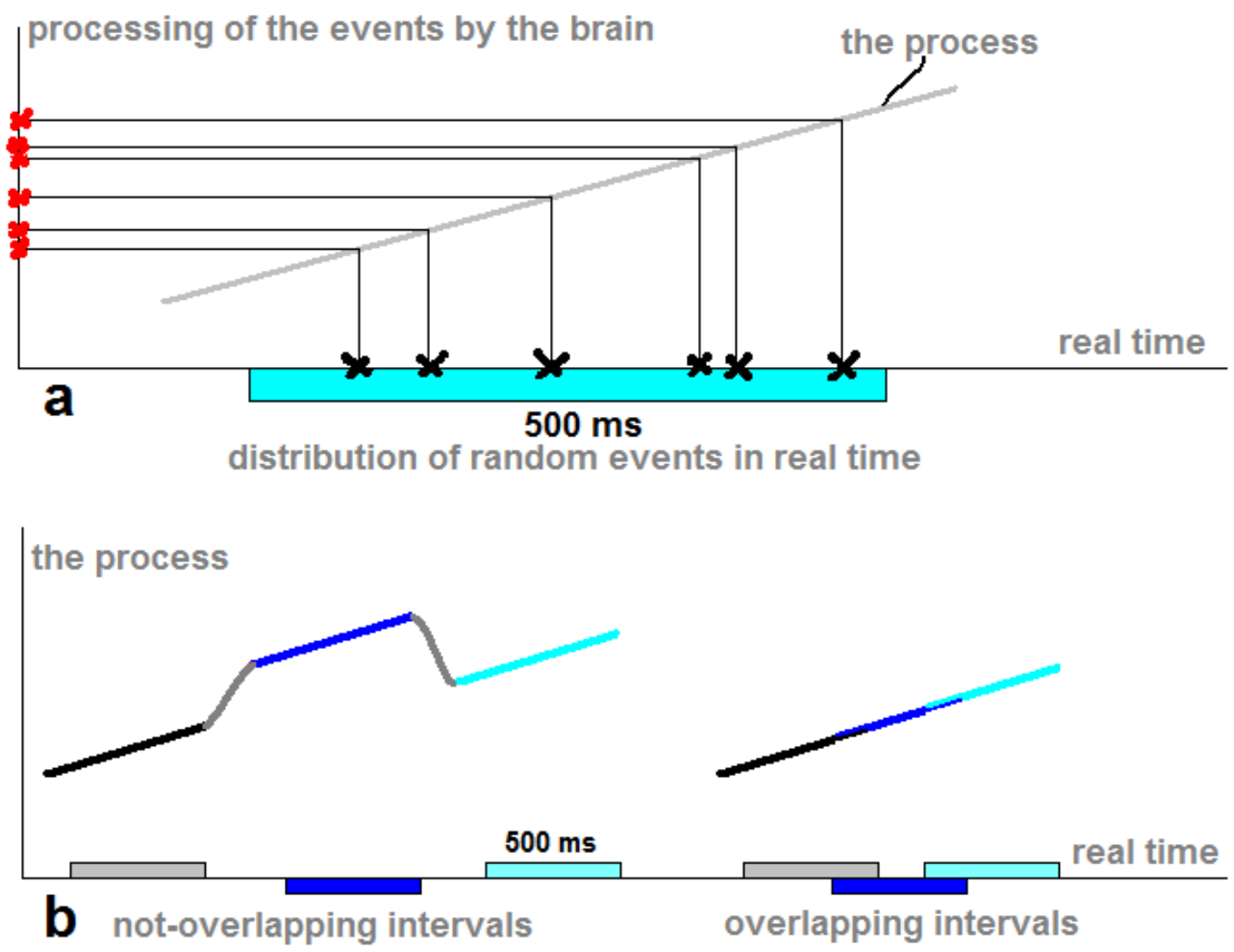

Fig.1. **a**. Temporal processing of the events by the brain must be similar to the actual distribution of the events in real time, at least within short time snippets. **b.** When the intervals of proportionality do not overlap, the process overall can be more complex. If the proportionality intervals overlap, the process overall is proportional, with the same slope everywhere.

Let us call the process and its interval measuring mechanism, the *Snippet Inner Time,* to emphasize that, though it must be defined and (almost) proportional to the real time during the whole period when the TIP activities are possible (which could be many hours of wakefulness), the measurement of only short time intervals (of few seconds or shorter) is required[4].

On the other hand, the Snippet Inner Time, SIT, must be limited, because the value of a scalar property of any internal process that can be implemented biologically is limited. But then it cannot be proportional to real time all the time we live (unless our life span is very short), Fig.2. Also most likely the SIT process that grows proportionally to time with increments of few milliseconds is quite complex and cannot decrease in an instant to renew its proportional task.

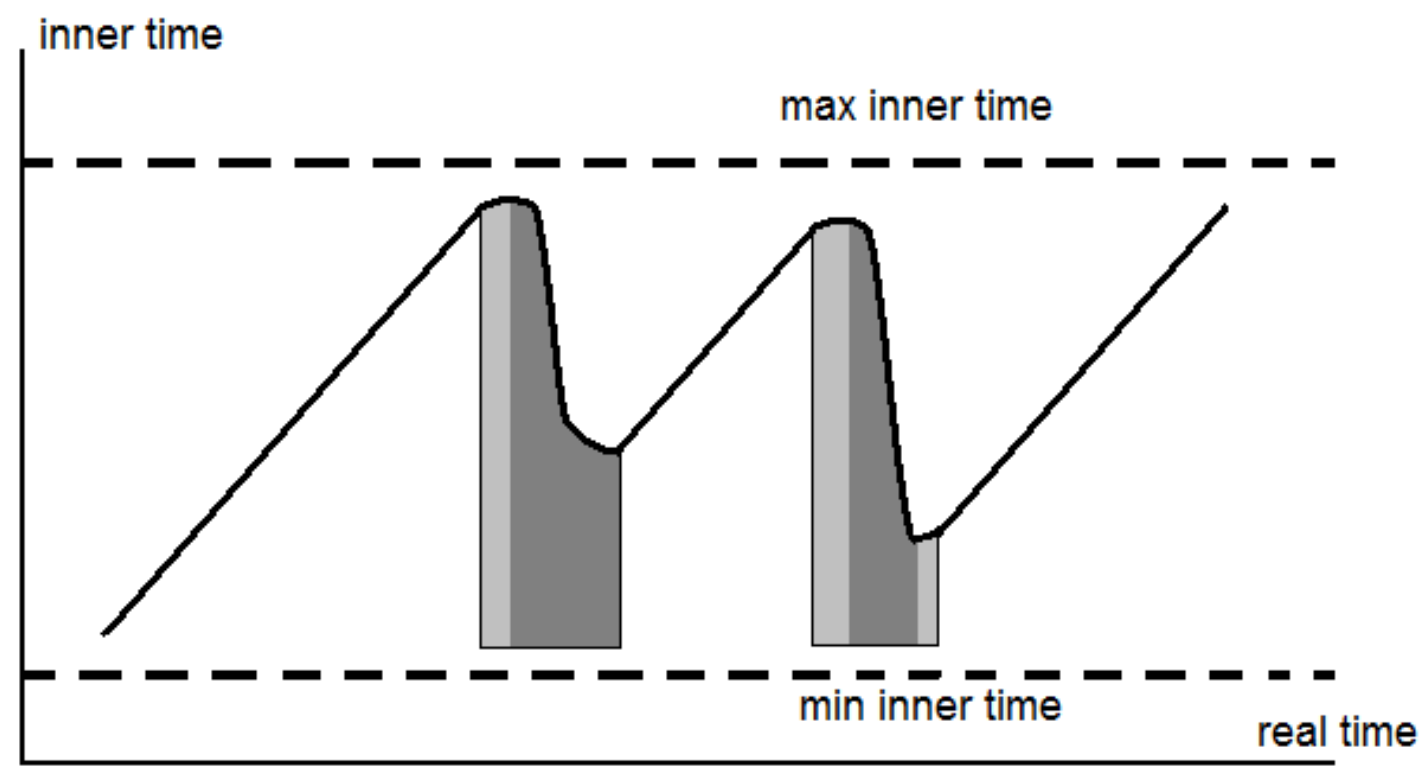

Fig.2. Snippet Inner Time (SIT).

During the periods of time when SIT is not proportional to real time (shaded in Fig.2) a correct temporal interpretation of the TIP events will not be possible and we will not be able to correctly interpret visual and audio input and to move normally using habitual motor patterns. Objectively these limitations would make the wakefulness period critically impaired, and as a species we would not exist if everyone had these limitations. However, both from subjective experience and from knowledge about the critical impairments of others (Helen Keller…), we know that wakefulness is not reduced to these three activities.

What does it mean to be awake? Let us assume that, subjectively, wakefulness is awareness of existence. Existence is existence in time, and thus, subjectively, wakefulness – at least indirectly – is awareness of time (perception of time). This requires some internal mechanism of time perception. The time perception mechanism is very different from SIT, in two ways. We know from experience that the equal periods of time might be perceived as very different depending on what we are doing, e.g. if we are bored or busy. In contrary, SIT must measure equal real time intervals as equal internal ones, so that we perceive the same words the same way as they are being spoken, are able to use habitual motor patterns etc. The other difference is that the range of subjective time perception is from several seconds to minutes and hours while SIT measures time intervals in the milliseconds range only.

---

[4] We draw the proportional piece of the graph as ascending for convenience.

Still let us hypothesize that these two mechanisms are related and that the perception of time is built on the top of “low level” mechanism of SIT, just as consciousness cannot function without lower level neuron activities. When SIT is taken off-line for resetting, which is periodically inevitable for normal wakefulness TIP activities, the perception of time also will be turned off. Then the light and dark shaded periods of the graph (Fig.2) indicate periods of drowsiness and sleeping, both objectively and subjectively.

Since the time-proportional Snippet Inner Time mechanism is necessary for our basic wakefulness activity, sleep is a “mathematical necessity”, because, mathematically, a limited internal process cannot be proportional to real time *all* the time.

## *Time-Bearing Oscillator Hypothesis*

The glymphatic brain waste removal system is by an order of magnitude more active during NREM sleep than at any other time [5]. If the principle cause of sleep is to reset the Snippet Inner Time mechanism then perhaps, studying what is discharged by the glymphatic system can lead to the discovery of what is the biochemical mechanism of SIT. The materials that are expeditiously removed by the glymphatic system during NREM sleep must have been gradually accumulating during wakefulness and may be the components of the SIT clock mechanism. We are not aware of any such research as yet.

Meanwhile let us imagine a process that might be theoretically possible in the brain and that could serve as a SIT clock. Snippet Inner Time should be about proportional to real time for 12-16 hours a day, and there should exist a mechanism that distinguishes and measures a few millisecond real time intervals from the change of the SIT value from now, so that the TIP activities can function.

Suppose there is a *variable frequency* oscillator circuit in the brain, the frequency of which changes over time monotonically. Let us call it a “time-bearing” oscillator, TBO. Suppose that when each TIP event is recorded a “tuning fork” oscillator is added to the record as the event time-tag. For the tuning fork, TF, the *natural frequency is fixed* and set to be equal to the TBO frequency at the moment of the event arrival. Then the natural frequencies of the recorded TFs will differ from the current frequency of the “time-bearing” oscillator, TBO; the earlier TF was set, the larger the difference. If the oscillation of TBO excites TF, it will create a forced oscillation in the TF. The amplitude of the forced oscillation, compared to the excitation amplitude, can serve as the measure of the time interval between now, the present moment, and the time of the event; the larger the amplitude, the smaller the time interval.

Qualitatively, the ability of such a mechanism to measure short time intervals between current and recent events can be seen from Fig. 3. A very small change of the TBO excitation frequency from the TF natural frequency will lead to a significant change in the forced oscillation amplitude, if the TF damping is small. The latter amplitude might be a very sensitive measure of short time intervals.

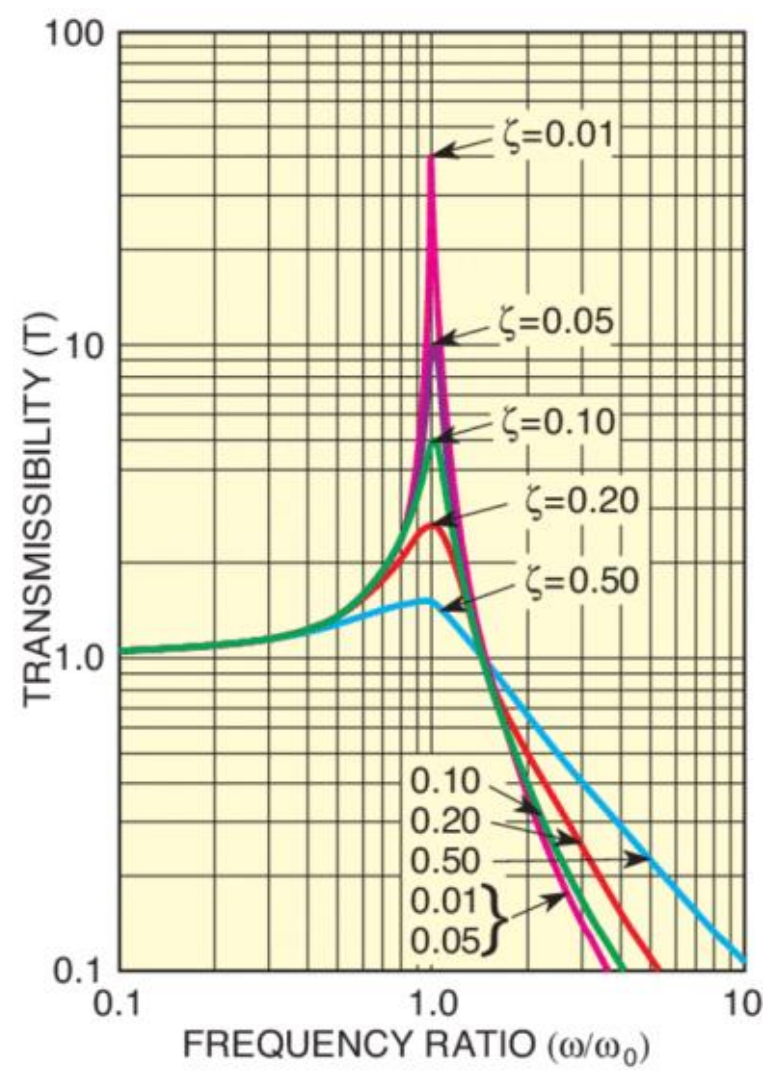

Fig.3. Dependence of the amplitude of the forced oscillation on the damping parameter ζ; $\omega_0$ – natural frequency; ω – excitation frequency.
https://www.newport.com/medias/sys_master/images/images/h1c/h0c/8798119198750/Fig-D-Transmiss-600w.gif

Suppose that the time-bearing oscillator frequency changes more or less uniformly from 4 Hz to 8 Hz over 16 hours[5] and the TF damping parameter ζ is around 1e-6. It is easy to assess [Appendix 2] that, close to the TFs natural frequencies, time intervals of 10 ms will change the forced oscillation amplitudes multiple times, which is detectable. Then monotonic ("almost" uniform, see below about variations of SIT) changing frequency of the time-bearing oscillator has a potential to be the Snippet Inner Time. The resonance-based mechanism of measuring SIT intervals is "reasonably natural" as it does not need a counter. It might be very efficient in sorting the recent events by their time interval from now, as the events can be sorted by the strength of the echo to TBO excitation.

Can an oscillator, such as the TBO, exist in the brain, having a frequency that changes over time at about the same rate monotonically? It is already discovered in rats, where it is found that the firing frequency of cortical neurons increases monotonically during wakefulness [6].

Let us speculate a little about the possible neuronal mechanism of the variable frequency oscillator. It is known that the number of $A_1$ adenosine receptors, $A_1$ARs, available for binding increases during wakefulness and decreases when asleep [e.g. 7-9]. When adenosine binds to its receptors, neural activity slows down, and we feel sleepy. Can the number of the available $A_1$ARs control the time-bearing oscillator frequency? Can binding/unbinding of $A_1$ARs change the frequency of some oscillation loop in the brain, similar to opening and closing the flute holes which changes its tone?

[5] The frequency range in this example is picked because EEG appears to be the most dependent on the waking duration in this range [Appendix 2]). However, if TBOs exist, theirs parameters might be different, depending on the species and individuum.

Using the flute analogy, how many $A_1AR$ holes should the adenosine controlled flute have, given that about 1e+7 different tones are needed to have a new tone for each 10 ms interval during the 12-16 hours of wakefulness? For the real flute, just 22 holes would be more than enough. Of course, it is unlikely that a set of brain receptors would use all the possible combinations of their states, active/inactive, nearly as efficiently as it is done in the binary representation of a number. But there are trillions of receptors in the brain, and the number of $A_1ARs$ that could be involved in the brain time-bearing oscillation loop, "the adenosine flute", could be significant.

Of course, referring here specifically to $A_1$ adenosine receptors is quite arbitrary. There are many types and subtypes of other receptors ($GABA_A$, serotonin, histaminergic …) that are involved in the wake-sleep dynamics [10]. What we are trying to say is that a biological implementation of TBO does not look impossible a priori, because the existence of the many different oscillations is known, and whatever controls these oscillations may have billions of different states. Even if during wakefulness all the TBO controlling receptors switch their states one way and during sleep – the other way (no two-way switching within waking or sleeping), even then, just having 1e+7 receptors to control the TBO loop would be enough. We will keep referring to this supposed implementation as "the adenosine flute" simply for convenience, and we'll continue the analogy by saying that at the beginning of the normal wakefulness cycle all the flute holes combinations are open (clean), then during wakefulness they become closed (dirty), one by one, which changes the flute tone (the TBO frequency). During sleep, conversely, the flute holes combinations become clean again, which rewinds the Snippet Inner Time clock.

## *Variations of the Snippet Inner Time Slope*

The slope of the SIT on the graph (Fig.2) has to be constant during wakefulness so that equal real time intervals are processed internally as equal and thus the TIP experience learned at one time is applicable at another. Yet it is hardly possible that a biological process inside us can change at *exactly* a constant rate within a several hour time span. This unavoidable variation may not matter, within some limits, as e.g., we do recognize the same voice and the same words as the same in the morning as in the afternoon. There must exist a recognition mechanism that provides "temporal invariance" within some SIT slope change range, similar to how within some limits we recognize tilted images as identical to those that are upright.

On the other hand we know that in some situations our movements become awkward, e.g. when we are sleepy or under the influence of alcohol. We tend to overreach, e.g., we reach for a cup of coffee and may overturn it. This may be explained by a significant variability in the SIT slope, Fig.4. Suppose that the "muscle memory" (the SIT-motor sequence) needed to grab a cup was learned with the SIT slopes at Fig.4a. If the SIT slope becomes much smaller, Fig.4c, then the learned sequence of SIT intervals will match much longer real time intervals. The muscle will work longer than is needed to perform the task and we overreach.

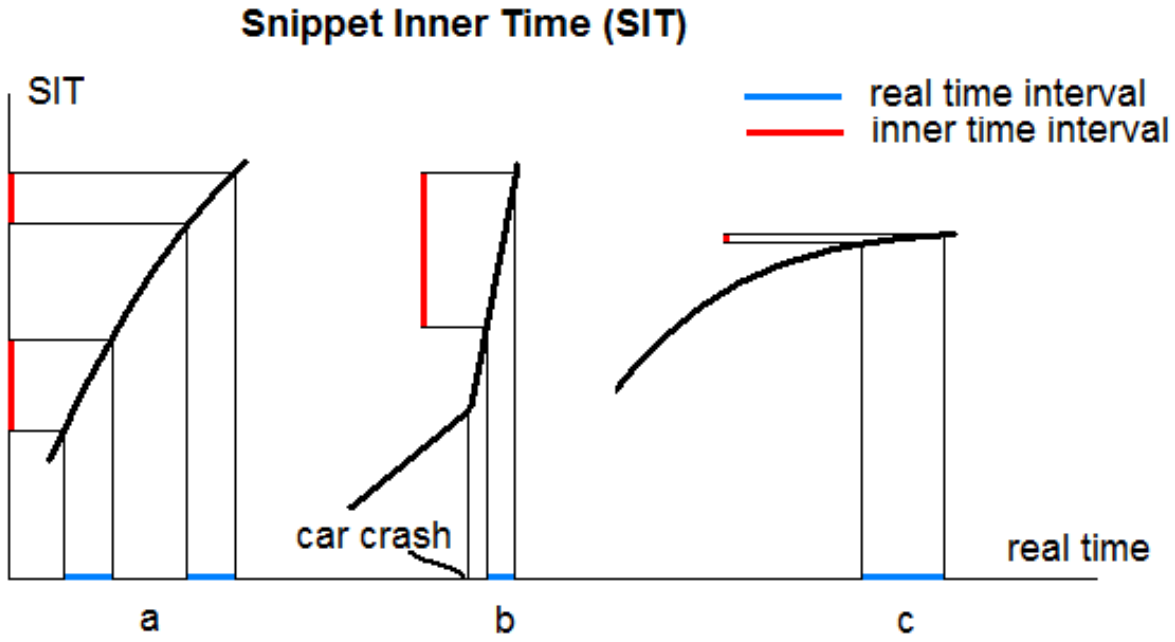

Fig.4. a. Equal real time intervals will be processed as different inner time intervals if the slope of SIT changes; b. If the slope increases by much, a short real time interval may be perceived as a long one; c. If the slope decreases by much, different real time events will blur into one.

Why, when we are well awake, is it easy to catch a ball, but when we are sleepy we miss? The same real time intervals will be processed as different ones, if the slope of SIT graph changes. If the slope drops to zero, the outer world events that are different in time will blur into one and we will fall asleep, Fig.4c. On the contrary, we may perceive[6] the outer world in slow motion, which would occur if the slope of the curve increases significantly, Fig.4b. There are multiple observations (e.g. [45]), including the author's personal car crash "experiment" [Appendix 1], that this may be happening under certain critical circumstances. These observations are questioned in [11], where based on the "free fall" experiments they concluded that the time slowing down effect is the function of recollection. But as it is shown in [Appendix 1] that conclusion is not reliable.

A smaller increase of the SIT slope, caused by a perception of a threat, will explain why an object that is moving towards the viewer may be perceived as moving longer (and hence, slower) than it does (e.g. [44]). It would be interesting to know if observing a looming object will simultaneously affect the hearing perception. E.g. in [44] experiments the steady, looming and receding disks were observed for 500 ms, which is about the same time as the duration of the word "one". Would the perception of the same recording of the word "one" depend on observing the moving disk? Such an experiment is not difficult to arrange, and the result could clarify the link between the timing mechanisms behind vision and hearing. We are not aware about such experiments yet (and cannot arrange them).

It is also known that there is a significant difference between the EEG of active and quiet waking. This may be is related to the expectation that, from the evolutionary perspective, during the quiet waking (closed eyes, relaxed) the SIT slope should drop, saving the SIT resource for the later.

---

[6] Perception is much slower and subjective a thing than our ability to act. The examples are endless, in particularly in sport activities, where after some training we move much faster than we can think. Or, when we are typing on a keyboard, do we really perceive each letter we type? After some training typing is mindless, often not only literally.

## *Snippet Inner Time vs the Circadian Clock*

The circadian oscillations are observed across the living nature and often are referred to as a circadian inner "clock"[7]. Our sleep-wake pattern also adheres, in general, to the 24-hour cycle, but there is no immediate dependency between sleep-wake and the circadian oscillations of our various inner functions. The latter are, on the cellular level, generated by the 24-hour cycling of specific proteins and need much time for adjustment, e.g. if we travel and experience "jet lag". In contrary, we can be woken up on a dime, and might fall asleep in the middle of the day. The circadian oscillations create a sleep or wake drive, but do not control the sleep-wake status.

On the level of the brain activity, the sleep-wake homeostasis is often presented as "Process S", Fig.5. This is similar to the Snippet Inner Time graph on Fig.2.

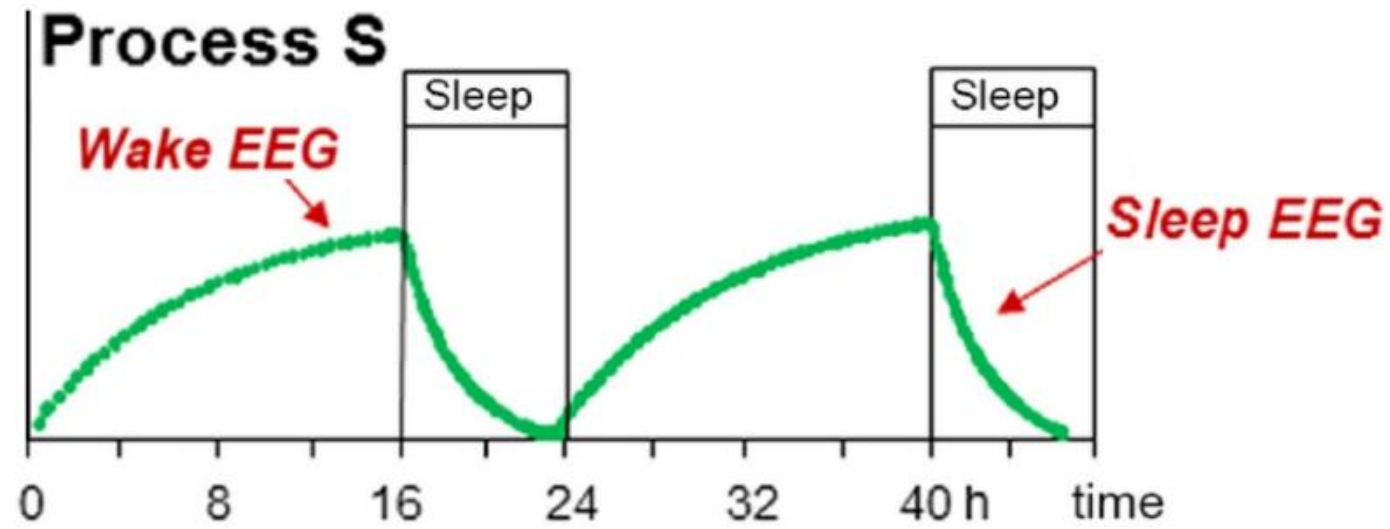

Fig.5. Process S (sleep drive; sleep-wake homeostasis), [12].

The similarity of the graphs is natural, because both represent the wake/sleep cycle (and because any sine type graphs on close periods would look alike). But there are two principal differences.

First, the Process S is a phenomenological model that generalizes a wide spectrum of the experimental research, but does not explain why it happens this way. The Snippet Inner Time model says that that is the only way it can be; there is no way around it. It is logically deduced as the only way that some very critical wakefulness abilities can function.

Second, the Process S value, the sleep drive (the graph ordinate), represents in a general way, characteristics that are integral to our state. In distinct to this, the slope (derivative) of the ascending branch of SIT graph reflects the rate at which we operate at any single moment of wakefulness, at a millisecond level, with the outer world and within ourselves. At the same time the descending SIT branch tells us nothing except that the SIT is being reset ("rewound") to some smaller value – as small as possible during a good sleep – that will be used as a starting point for the next coming wakefulness. There is no known logical reason why the time intervals should be measured during sleep; we drew the descending branch just to fill the void and show that SIT cannot be reset instantly, because there is a quasi-continuous[8] biophysical process behind it. Maybe it would be better not to draw it at all not until the actual mechanism of SIT implementation is discovered and we know if there is something temporal to measure during sleep.

---

[7] A real clock should have a time interval measuring mechanism. Nothing is known about the existence of such an internal mechanism that would be related to the circadian oscillations.

[8] Quasi-continuous is the sense that it cannot change much instantaneously.

## *Yawning hypothesis*

Wakefulness requires SIT to be growing almost proportionally to real time. SIT, as any biologically possible process, is bounded. Then, when it reaches its upper limit there is no work around and we *have* to fall asleep. Note that we might fall asleep well before SIT reaches its upper limit, because there is nothing to do and we are bored, and evolution should have found a way to use this time to set us ready for future battles. But now let us talk about the opposite situation where something very important is happening at this very moment, something very interesting or life threatening that we do not want or cannot afford to miss, but the SIT is reaching its upper limit. If it gets there, we fall asleep and miss the action, or even die. Should evolution not have found some defense tool to prolong wakefulness for a little bit, to pass through this important time?

Let us speculate that *yawning* is the wakefulness short-time extension tool, allowing the SIT levels to drop a little during the couple of seconds of yawning, which would allow us to keep going for another several minutes before reaching the SIT upper limit, Fig.6.

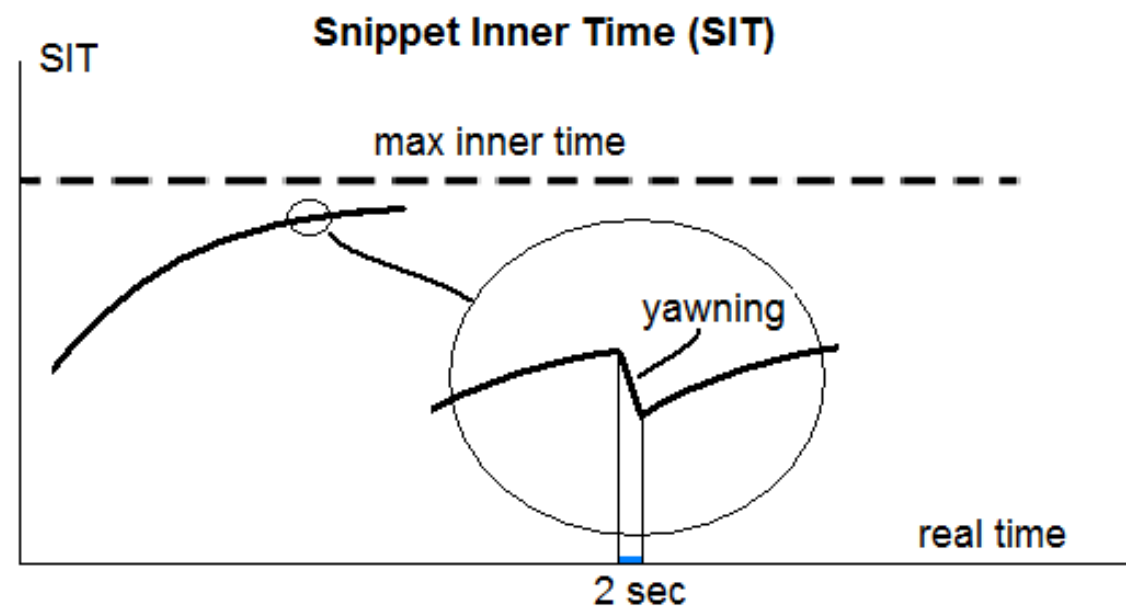

Fig.6. Snippet Inner Time drops a little during yawn, as if during a very short sleep.

Table 1 summarizes the consensual facts about yawning and explanations of the facts by the SIT dropping (yawning bluff) hypothesis.[9]

Now let us review if the yawning bluff hypothesis contradicts the experimental study that came to the following "*Conclusions: The data therefore do not support an arousing effect of yawning or a role in regulation of vigilance*" [14]. But does this contravene the Fig.6 yawning graph? In the research, they assess arousal by comparing the thirty-second samples of the brain waves (EEG) before and after yawns. Other ways of assessing arousal were reviewed. They found no difference before and after yawns, and came to the quoted conclusion. But if we look at the SIT graph pieces, Fig.6, before and after the yawn, the arousal levels (which are derivatives of the graph) are *expected* to be almost the same, only shifted in time. The SIT yawn-bluff hypothesis does not suggest that after the yawn we become more aroused; it just suggests that yawning delays the onset of sleep.

[9] Of course, if the yawning bluff hypothesis is correct then the next question is, how is the SIT drop related to the mouth opening while yawning? There may be different speculations about the mechanism. E.g., it is known that hypoxia increases the adenosine level, which also occurs during waking. Could it be that the influx of oxygen during yawn works the opposite way, as in sleep? We favorite another hypothesis, based on what we know now, that stems from the observation that mouth opening during yawn is rather strained. The state of unconsciousness is similar to the NREM sleep (though there are some known EEG differences). A person can be knocked out into unconsciousness by a punch into the jaw, presumably, because the jaw mashes into the brainstem and glossopharyngeal nerve clusters. Then could it be that yawning works as a micro-punch, massaging these nerve clusters and leading to a micro-knockdown – SIT drop?

Table 1. Yawning hypothesis.

| # | Consensual experimental facts | Explanation by yawning bluff[10] (SIT drop) hypothesis |
|---|---|---|
| 1 | We yawn when we are drowsy but something causes us to delay going to sleep. | Yawning delays the onset of sleep, because SIT drops a little from the SIT upper limit, Fig.6. |
| 2 | Yawning "*is also found under fearful and stressful situations, as was commonly observed during the First World War, in troops waiting in the trenches for the 'whistle to blow' and the order to 'go over the top' and confront the enemy. It is also seen in athletes before a competitive race, with actors before going on stage, and in migraine sufferers just before an attack"* [13]. | If we are expecting that something critical is about to happen would we not want our action-controlling inner clock, the SIT clock, to be "rewound" in full? To do this we usually try to get a good sleep before an important event. In any case, a drop in SIT levels at the last moment, even a small one, could be essential. |
| 3 | Yawning is contagious; seeing others yawning may provoke us to yawn. | This is directly derived from 2, given the fact that as social animals we take cues from others about what is going to happen. If they are yawning then maybe something important is about to happen, so I better get ready too. |
| 4 | For many of us, when we yawn, our body would shake, especially upper body and arms. | A few seconds yawn is not long enough to turn around the whole complex central nervous system, so we do not fall asleep and fall on the ground all at once. But if during the yawn the SIT slope reverses then the normal "muscle memory" based control will fail. The antagonistic muscle pairs might produce unbalanced forces and result in an unwanted movement. If this is noticed and corrected with a delay the result will be the body part oscillation. In general, the muscles will be moving controlled only by a much longer and slower feedback loop than the "muscle memory" provides; and a slow feedback in a dynamical system often results in oscillations. |
| 5 | If something is said to us when we yawn we miss what it was. | If the temporal sorting is still turned on when SIT is decreasing, the incoming sounds will unfold inside our brain in the reverse order and we'll hear the words backwards. If the temporal sorting is turned off during the yawn then all the sounds will be jumbled together. In both cases we will not understand what was said. This explanation suggests that during the yawn we do hear the sounds, we just cannot sort out what they mean.[11] |

[10] A "bluff" here means a "cliff", because of the shape of the Fig.6 graph.

[11] There is an explanation of the yawning "deafness" that it is because "*The act of yawning activates the tensor tympani muscle. Activation of the muscle dampens hearing, leading to perceived temporary deafness."* This maybe is true (though we are not really deaf during the yawn, we know that something was said, but what was it?); we suggest an alternative explanation that comes from the SIT bluff hypothesis.

It is interesting that in the same research that comes to the quoted conclusion they refer to another research study where it was concluded that within 10 to 20 sec near-yawn intervals there may be transient changes in the brain waves typical of a mix of wakefulness and shallow sleeping *(transient decreases in Delta and increases in Theta, Sigma (spindle), and Beta frequencies associated with yawning*) [15]. Perhaps this confirms that yawning does what a very short sleep would do, which according to the yawn-bluff hypothesis is to rewind SIT a little.

## *Sleeping hypothesis*

If the main (and only??) purpose of sleeping is to rewind the SIT clock then why does sleeping appear to be so complex, with different stages, paradoxical rapid eye moment (REM) sleep, several non-rapid eye moment (NREM) stages of sleep, and – most of all – with the amazing sweet dreams and dreadful nightmares?

Let us think about sleep from the point of view of the following wakefulness. When we wake up, sometimes we are fresh and ready to go, feeling that we had a good sleep. But often we are sleepy, we feel not so good and we need to sleep more. We know very well why the latter happens, it is because something woke us up before we were ready. It could be a phone-call in the middle of the night, a toilet call or a (tactless) interference of the internal *circadian clock* that ignored that we went to bed very late yesterday. It could be a disturbing dream.

But what wakes us up after we have *had* a good sleep and we are absolutely *ready* for the full-fledged awakening?

Sometimes it might be a case when the circadian clock performed in sync with our actual (nowadays, often hectic) life schedule. But we do know that if we go to bed earlier than usual and manage to fall asleep we might get up in the middle of the night because we just don't want to sleep anymore – and this would be well before the circadian clock wakes us up. What wakes us up in this case?

How would the brain know that we have *got* enough sleep, that it *is* ready to start the wakefulness run without the risk of breaking a leg or setting the car to "D" instead or "R" when backing out of the garage?

Table 2. "REM sleep is a test-run of the brain readiness for the wakefulness" hypothesis.

| # | **Consensual experimental facts** [16] | **"REM sleep is wake readiness test-run hypothesis" explanation** |
|---|---|---|
| 1 | *EEG readings of brain waves in persons who are awake are similar to those in REM sleep.* | System testing must be similar to the actual usage. |
| 2 | *Neuronal activity is high during REM sleep, brain blood flow and oxygen use are actually higher during REM sleep than during intense mental or physical activity while awake. Heart rate, respiratory rate, and blood pressure increase during REM sleep*. | Usually during testing the system is forced to endure a higher load than is expected during the regular use, because this adds to the system reliability |
| 3 | *Most somatic motor neurons are inhibited during REM sleep, which causes a significant decrease in muscle tone and even paralyzes the skeletal muscles.* | This is necessary for safe testing. For example, when we test a car in a garage, we would set the stick to "Neutral" (clutch off), and then we may run the engine, make it roar, test the lights etc. |
| 4 | *The main exceptions to this inhibition are those somatic motor neurons that govern breathing and eye movements*. | To continue the car analogy – while the car is in neutral, we may let the fan, a/c and the windshield wipers run to test them. |
| 5 | *REM sleep is also the period when most dreaming occurs* | What is most amazing about dreams? They may bring up old forgotten memories and they may mix together some completely unrelated facts and experiences. In general, dreams may create any mixture of images and experiences, any mess, and keep processing it, as if it makes sense. Isn't this what a good software tester does? Any software development manager knows that a good tester will keep trying different possible input combinations, bizarre as they may seem, to make sure that the software will still keep running even with the bizarre input and that at least it won't crash. So, is dreaming just a way of testing our ability to accommodate whatever input combinations brain can assemble – however bizarre – (and not go berserk)? |

Our brain is a very complex system, most likely the most complex there is around. What is the only known way to get at least some assurance that a complex system will run properly after it is started (e.g. after a maintenance)? The only known answer is to *test* the system before starting its regular run, in as safe an environment as possible.

Then it is natural to hypothesize that sleep must include two phases, the phase of the system restoring/cleaning (sleep homeostasis), which definitely occurs during sleep, and the phase of testing if the system is clean enough (sufficiently restored). These phases may alternate, because if testing is showing that the system is not clean enough it must go back to cleaning.

We hypothesize, based on the reasons that are obvious from the Table 2, that REM sleep is the phase of testing.

This hypothesis is also in agreement with the typical distribution of REM sleep over the sleep cycle for a healthy human that is shown on Fig.7

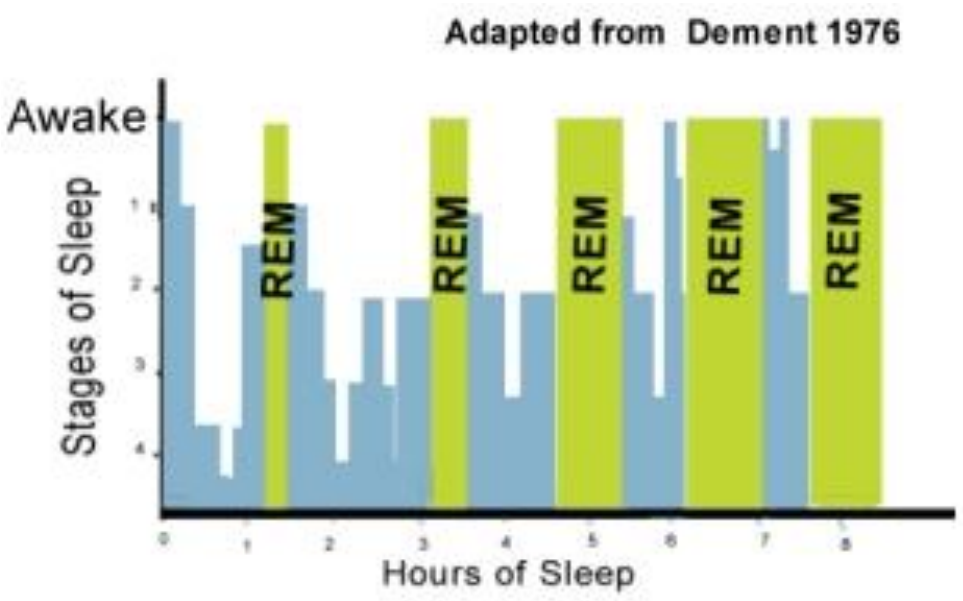

Fig.7. REM intervals become longer, while NREM shorten during the sleep cycle [17].

If REM testing of the system for cleanness were to find that it is not clean enough, will the brain keep testing, or will it get back to cleaning? Continuing testing would be a waste of time and resources, so the brain will get back to cleaning. Then, after a while, when perhaps the system has become clean enough, REM testing will start again. This time it will take longer to find the dirty spots, because the system has become cleaner. The cleaning periods, in turn, will become shorter because as the system becomes cleaner, the brain will be able to return to testing more quickly. The longer the cycle, the more testing is carried out and the less fixing remains to be done, which is pretty typical for the software developmental cycle. Finally, the product is released, or you wake up.

So, if the interpretation of REM sleep as testing of the brain’s readiness for wakefulness activity does not comply with the known facts, then what other interpretation would fit?

Now let us turn to NREM sleep, which reverses the biochemical changes of wakefulness, provides sleep-wake homeostasis, and, in our analogy, cleans the “adenosine flute” (the temporal mechanisms of the brain). NREM sleep has specific brain waves (slow waves, spindles, K-complexes…). What is the task of these regular oscillations with large amplitude? There are 100s of researches and speculations about this, but let us start with a question: if we want to make something clean, is it enough to put it in a clean environment and wait until it becomes clean by itself, or do we apply additional efforts to make it happen faster? The human experience in this area is illustrated in [18]. In theory, we could put the cloth into clean water and wait for few months until it becomes clean, but that might be a little too long to wait!

Any brain receptor state is to some degree stable, otherwise it would not have been identified as a state. That stability means that in order to change the state a certain amount of effort must be exerted. Stepping back to the analogy of cleaning the adenosine flute, all the holes combinations need to be reopened in order to rewind the SIT clock. The question is how does this change in the state of the adenosine receptors (or any other receptors) come about? Is it enough to change the extracellular adenosine concentration and wait until more $A_1A$ receptors become available? Or would it be too long to wait?

When we look at the larger NREM sleep waves, their amplitude and regularity suggest that the work they do is not of a subtle, intricate nature but is a straightforward, overall task applied indiscriminately in bulk. Could it be that they are just energy excitations needed to expedite "the flute cleaning"? During wakefulness the Snippet Inner Time clock is needed to synchronize many different systems, sensory, motor, cognitive…, each consisting of many subsystems. Different mechanisms are needed to provide the synchronization. Resetting/cleaning all these mechanisms may require diverse efforts that may manifest as different wave types.

We hypothesize that NREM sleep provides temporal related cleaning in two ways. First, it resets the SIT clock, as low as possible within the available cleaning time. Second, the time-tags, associated with every event, become obsolete as the time span between the event-time and now, the present moment, increases. So one would expect that the time-tags gradually disintegrate as a useless data. However destruction of the time-tags also require energy and without an additional effort might go slowly. Since the time-tags are created as copies of SIT, the mechanism of their storage must be influenced by the same factors as the SIT mechanism. So we hypothesize that NREM also facilitates removal of the time-tags. Another reason why the time-tags must be removed after SIT is rewind is that, if the tags were not removed, it might lead to a false association (conditioning) between the events that occur before and after the sleep, when the SIT values were the same.

Good cleaning/resetting of a really complex system often requires that it is at least partially disassembled. So it might be that during NREM sleep some connections between the brain subsystems are "unplugged and cleaned" and that during REM sleep they are reassembled and tested.

So, could it be that the sleep stages are for "brainwashing" – literally – applied to the receptors, and for checking that everything is clean enough? The high activity of the glymphatic system during NREM sleep [5] may be another indication that this is so, as washing needs much drainage.

## *Discussion*

The suggested theory contains ideas of a different hypothetical character.

The existence of the activities that require accurate measuring of short time intervals, TIP activities, is a fact, obvious and trivial from the point of view of the post-Newtonian physics. This fact usually is not discussed explicitly, though each research that observes the ability of an animal to generate highly reproducible temporal sequences implicitly confirms this. As an example, the research [19] observes a highly reproducible sequence of the hippocampal cells firing when a rat runs along the same 15 sec long route.[12]

Meanwhile, as it is noted in the fascinating review *The inner experience of time* [20], "the striking diversity of psychological and neurophysiological models of „time perception" characterizes the debate on how and where in the brain time is processed". Despite the diversity, until this our research not a single concrete internal process has been explicitly identified that cannot function without an inner time, aside from the generic understanding that "the perception of time is an essential and inextricable component of everyday experience". This probably is a major factor why "no conclusive answers to the questions of which neural substrates and what kind of neurophysiological processes could account for the experience of duration have been established", because without a specific understanding where the internal time is used it is not clear how and where to search for its physiological mechanism.

The *constant (absolute)* time-tags is a very plausible hypothesis, for the incoming from the outer world events (stimuli), because the constant time-tags appear to be simpler and more efficient than the *variable (relative)* time-tags, when it is not known which events are related to each other. If the constant time-tags hypothesis is true then SIT mechanism of measuring of short time intervals is logically deducted from it. If SIT *is* the mechanism of short time intervals assessment, then periodical sleeping becomes a logical necessity. A comparison can be made with the modern computers that cannot function without a system clock that generates system time. The difference from the putative animal SIT is that for a computer the range of the system time is 100 – 10,000 years, depending on the computer system. (Getting the computer system time out of range would lead to a chaos in the data/file management. That was expected – and avoided – when "Y2K bugs" were fixed at the end of the previous millennium.)

In contrary, for the events generated by an animal (e.g. replay of the muscle memory), the time-tags, obviously, are relative (e.g. to the start of the events chain). This means that depending on the character of the event different types of the time-tags might be used.

The TBO realization of the SIT mechanism initially looked a bizarre and pretty speculative hypothesis. It did not originate from any experimental observations; instead, it came up from a discussion with Marat Rvachev, who argued that resonance might be very important in the brain activity. However, we kept exploring this TBO hypothesis for two reasons. First, if TBO exists then its oscillation might be observable by the existing measuring equipment. Second, however bizarre it might appear, the oscillators with frequencies that change over time monotonically might naturally exist, influencing evolution of the organic matter. If a closed shell filled with gas is placed at a fixed depth in the ocean then the ocean tides will periodically change the shape of the shell, making it a possible variable frequency mechanical resonator. If the shell is filled with liquid and bounded by a filtering membrane then the changing tidal pressure might change different ions concentrations inside the shell and make it a potential variable frequency bioelectrical resonator.

---

[12] Amazingly, the hippocampal cells firing sequence is, in a compressed manner (~60 times faster), preplayed before and replayed the after the run. The preplay is in the same sequence order, while during the replay the sequence order is reversed. Speculating on this, let us add here one more hypothesis. Suppose that the memory of a temporal sequence is organized in the brain as a stack memory (which is known in the computer science as a simplest and fastest way of the sequential storage). Then the observed preplay before the run may be extraction of the memory from the stack, and the replay after the run may be storage of the updated memory back into the stack, for the future use. (The storage in the stack, obviously, always is in the reverse order vs the data usage.) The observed time-scaling may be due to the memory handling time.

Only much later we ran into the discovery by Vladyslav V.Vyazovskiy and his team [6] that there really are oscillations in the brain with the frequencies that change monotonically over the course of wakefulness, Fig.8.

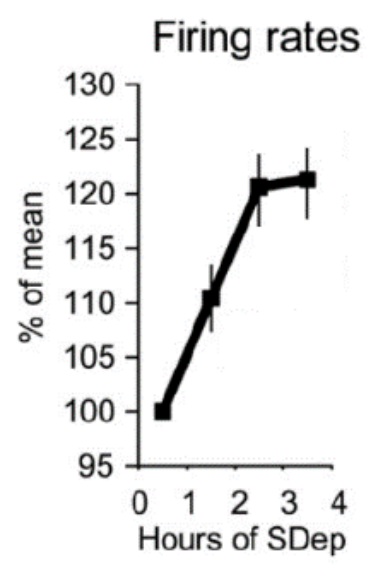

Fig.8. Cortical neurons firing frequencies change in rats during 4 hours of wakefulness (relative to the 1st hour) [6].

Of course, by itself the existence of the neuron oscillator that changes its frequency monotonically during wakefulness does not prove that its current frequency is attached to the event memory as the time-tag, in the form of TF, and the oscillator is TBO. However, a time-tag *must* be attached to the event memory, so that events could be temporarily differentiated. The minimum complexity of storing the tuning fork (TF) description as a combination of the brain receptors states is the same as the complexity of storing of any description of the current time, because the minimum complexity is eventually defined only by the required temporal discretization. Would not it be likely that evolution had found a way to use the monotonic variable frequency oscillator – which, as we already know, exists – to work as TBO, because it provides a simple instantaneous tool to temporarily differentiate the events by the strength of their echoes to now (to TBO of the present moment), emphasizing recent events and attenuating old ones? What could be another purpose of such a variable frequency oscillator?

Hence we hypothesize that the inner time homeostasis, the necessity to reset TBO frequency – rewind the inner clock – and impossibility of the full-fledged wake functioning without the running inner clock, is the reason of sleep as the inhibited behavioral state.

The yawning and sleeping hypotheses are indirectly confirmed by the well-known consensual experimental observations as described above.

For the yawning, we could not find a meaningful yawning theory that would cover at least half of the listed above consensual experimental facts. It seems there is nothing to set our yawning hypothesis against off.

Before comparing the sleep hypothesis with other popular theories about sleep let us recall the current experimental limitations: fMRI – temporal resolution a few 100 milliseconds and spatial resolution of 100s of thousands of neurons in each voxel [21]. Other experimental methods of the brain research produce even much less localized information. Objectively very little is known about how the information is stored in the brain. The theories about "memory consolidation etc." during sleep are not based on the experiments that can be *uniquely* interpreted. If we don't know how the brain memory works, is it really possible to make definite conclusions of how it is consolidated?

Let us consider several examples of the experiments hypothesized to confirm that the sleep goal is "memory consolidation", and show that the same results can be interpreted as confirmation of the SIT theory in a much simpler and thus less speculative way.

In a very interesting research of the relational memory [22] "the participants initially learned five ''premise pairs'' (A>B, B>C, C>D, D>E, and E>F). Unknown to subjects, the pairs contained an embedded hierarchy (A>B>C>D>E>F). Following a delay of either 20 min, 12 hr (wake or sleep), or 24 hr, knowledge of the hierarchy was tested by examining inferential judgments for novel ''inference pairs'' (B>D, C>E, and B>E). Despite all groups achieving near-identical premise pair retention after the delay (all groups, >85%; the building blocks of the hierarchy), a striking dissociation was evident in the ability to make relational inference judgments: the 20-min group showed no evidence of inferential ability (52%), whereas the 12- and 24-hr groups displayed highly significant relational memory developments (inference ability of both groups, >75%). Moreover, if the 12-hr period contained sleep, an additional boost to relational memory was seen for the most distant inferential judgment (the B>E pair; sleep = 93%, wake = 69%)… Together, these findings demonstrate that human relational memory develops during offline time delays."

Let us provide an explanation of these experimental results of the relational memory development based on the SIT ideas. Each event always has a time-tag attached to it, because upfront it is not known if the event will be involved into a temporal chain of events/conditioning. When the participants studied the premise pairs they learnt more than the relations A>B or B>C. Instead, they learnt (subconsciously) that $A_{1234567}>B_{1234567}$ and $B_{5678901}>C_{5678901}$, where indices are the time-tags of the times of the learning. Since the actual values of the time-tags are unknown we put in arbitrary values, 1234567 and 5678901, just to underline that the tags are different, because the learning occurred at different times. The length of the time-tag, 7 digits, is taken from the assumption that all the absolute time-tags within one wake cycle must be different; the wake cycle lasts about 12-14 hours and the time-tags allow differentiation of the 10 ms time increments. Though the researches randomized the training session to avoid establishment of hidden and uncontrollable dependencies, it was in principle not possible to get rid of the time-tags. And, of course, the last two learnt inequalities do not infer that $A_{1234567}> C_{5678901}$, because $B_{1234567}$ and $B_{5678901}$ are different. This is why the testing that immediately followed the training did not produce the inferred dependence, A>C. Overtime, after the time-tags disappear, in particularly being deliberately removed during NREM sleep, the inference becomes trivial. This explanation is logically simple and far less speculative than references to vague "process of consolidation" and "continued modulation of recently acquired information".

The research [23] and related preceding studies are devoted to procedural (finger-tapping) experiments. They "have shown that a night of sleep can trigger significant performance improvements in speed and accuracy on a finger-tapping task, whereas equivalent periods of time during wake provided no significant benefit. Furthermore, these overnight learning gains correlated with the amount of stage-2 NREM sleep, particularly late in the night." In particularly, it is shown that the more complex the task, the greater speed improvements are demonstrated after overnight sleep.

The absolute time-tags – SIT hypotheses provides following explanation of these experimental results. Suppose that the trial participant is trained the finger tapping sequence, 1 – 2 – 2 – 4, by repeating this sequence 3 times. What is *actually* being learnt is much less clean; it is something like this, $1_{1234567} – 2_{1234568} – 2_{1234569} – 4_{1234570}$, the 1st time, then $1_{1235567} – 2_{1235568} – 2_{1235569} – 4_{1235570}$, the 2d time, and $1_{1236567} – 2_{1236568} – 2_{1236569} – 4_{1236571}$, the 3d time, if each tap takes

10 ms (last digit in the index); the time difference between tries is 10 s and during the last try, before tapping the 4th finger the participant was distracted and it took 20 ms to tap. Recalling of the newly learnt tapping sequence soon after the training would be slower than recalling a "pure" sequence without the tags, because the sequence with the tags is heavier than the pure sequence and hence is more difficult to extract. Recalling with the time-tags will be more error prone, because, contrary to the assumption that during the training the same sequence was repeated 3 times, in fact, until the time-tags are removed, they are 3 *different* sequences. It is mathematically trivial that the longer the sequence the more it will benefit from the time-tags removal. Since the time-tags are eradicated during NREM sleep, it explains the experimental results.

The research [24] explores if spatial memories are strengthened in the human hippocampus during Slow Wave Sleep (SWS, stages S3-S4 of NREM). The participants were trained to learn a route in a virtual town. It was known from the previous and reconfirmed in that study that post-training sleep lead to improvement in the route retrieval performance, comparing to the pre-sleep testing. The study found a positive correlation between hippocampal activity during SWS (using PET and rCBF) and the improvement of performance in route retrieval, while "no significant reactivation of the hippocampal formation was observed during post-training REM sleep in the present experiment". In line with our SIT – NREM-REM sleep hypotheses, this may indicate that the time-tags of the events are stored in hippocampus and observed hippocampal activity during SWS reflects time-tags removal, which, in turn, improves the performance.

Next, let us review the experimental researches of the odor – visuospatial memories of the card placement, [25, 26], that "support the hypothesis that once an odor has become associated as the context of learned object locations, reapplication of the odor during subsequent SWS acts as a context cue that reactivates the new memories and thereby boosts their consolidation."

Let us provide alternative interpretation of the results. The significant difference between the cited above relational memory development results, the procedural memory finger-tapping skill and the special route learning vs the visuospatial memories explored in [25, 26] is that for the first three cited researches the results after sleep are *better* than before sleep. In contrary, the visuospatial memories in [25, 26] after sleep are *worse* than before the sleep, though *less worse* if the odor is present during training and following SWS. Why the results after sleep are better in the first three cases but not in the last one? Is it because in some cases sleep provides "additional training" (or "consolidation"/copying of information) and in the other cases it does not? Or is it because in some cases sleep provides sufficient removal of the unrelated to the intended experimental learning (like the time-tags) and in the other cases, for some reasons, this cleaning does not occur, or is not good enough. The second explanation is less speculative, and deletion of information (cleaning) is much simpler a process than creation a new one (including creation a copy).

To that end, it is of note that in the discussed olfactory experiments "the odor was always presented in an alternating mode of 30-s on, 30-s off periods to prevent habituation". Habituation is learning, and preventing habituation is also learning. Does the shift of the phase of the odor oscillation relative to the exposition to the visuospatial training patterns affect the result? Was it controlled in the experiment and is it even possible to study this, given the fact that the odor remains in the mask for a while after it was turned off?

The process of learning is collecting a puzzle from pieces that might be initially dirty by the data irrelevant to the intended *final*[13] learning result, where the dirt prevents the proper pieces to click together. The sources of the dirt are manifold, as the brain does not know upfront the intended result, what is and what is not the dirt. One of the inevitable sources of the dirt might be the time-tags that are always present for an event that may become a piece (stimulus) of trace conditioning, but might be a dirt relative to the intended learning. The time-tag dirt is being removed during sleep (SWS) in the relational memory, procedural memory finger-tapping and route learning experiments discussed above, but for some reason the removal is hampered for the visuospatial card placement tests in [25, 26]. It seems that conditioning of SWS with odor facilitates cleaning some of the dirt from the card placement memory. Comparing to this, the explanation suggested in [26] that "synchronized dialogue between thalamocortical and hippocampal circuitry and which eventually leads to the transfer of the representations to neocortical regions for long-term storage" is by far more arguable, where synchronized activity is an experimentally observed fact and the rest is a speculation.

In this context it is not possible to avoid a generic discussion of the "The two-stage model of memory consolidation". Here is a full quote from [27] describing the model as of 2010: "A key issue of long-term memory formation, the so-called stability–plasticity dilemma, is the problem of how the brain's neuronal networks can acquire new information (plasticity) without overriding older knowledge (stability). Many aspects of events experienced during waking represent unique and irrelevant information that does not need to be stored long term. The two-stage model of memory offers a widely accepted solution to this dilemma. The model assumes two separate memory stores: one store allows learning at a fast rate and serves as an intermediate buffer that holds the information only temporarily; the other store learns at a slower rate and serves as the long-term store. Initially, new events are encoded in parallel in both stores. In subsequent periods of consolidation, the newly encoded memory traces are repeatedly re-activated in the fast-learning store, which drives concurrent re-activation in the slow-learning store, and thereby new memories become gradually redistributed such that representations in the slow-learning, long-term store are strengthened. Through the repeated re-activation of new memories, in conjunction with related and similar older memories, the fast-learning store acts like an internal 'trainer' of the slow-learning store to gradually adapt the new memories to the pre-existing network of long-term memories. This process also promotes the extraction of invariant repeating features from the new memories. As both stores are used for encoding information, in order to prevent interference, the re-activation and redistribution of memories take place off-line (during sleep) when no encoding occurs. Because in this model consolidation involves the redistribution of representations between different neuronal systems that is, the fast- and slow-learning stores, it has been termed 'system consolidation'. For declarative memories, the fast- and slow-learning stores are represented by the hippocampus and neocortex, respectively."

The described model is an elaboration of the original hypothesis that the acquired information arrives first to hippocampus and then later on is redistributed to neocortex. The latter hypothesis stems from the clinical observations that patients with a damage to hippocampus cannot make new, long-term memories. This hypothesis had to be modified years later, after fMRI experiments have shown that hippocampal-neocortical activation and crosstalk occurs during (and continues after) initial information learning. Still, there are experimental results that "question the standard model of hippocampal–neocortical interactions" [28]. The latter results have a better explanation that "newly learned memory traces are integrated into neocortical regions during co-activation, while erasing recently learned information in the hippocampus at the same time". This explanation is well in line with our hypothesis, detailed below, that the absolute time-tags are being removed after they facilitated placement of the newly acquired memory in the context of the preexisting knowledge. Also it is of note that the quoted above statement, "As both stores are used for encoding information, in order to prevent interference, the re-activation and redistribution of memories take place off-line (during sleep) when no encoding occurs", may be not accurate because the long term memories can be acquired without sleep.

[13] But maybe relevant to the *process* of learning, see below.

Let us provide alternative, and simpler, interpretation of the same experimental results. When an information in being stored in a complex system, some attributes must be used for the information to be stored in a retrievable, and therefore usable, manner. The information must be stored in the context of the existing informational structure. Many of us, if not all, have an experience when we thoughtlessly saved a file in a computer, without paying attention to the file name and the folder in which it was saved. We know that it is very difficult, almost impossible, to find the file later on, though the information it contains may be important and relevant to some other information that is stored in the computer.

So let us hypothesize that each event that arrives into the brain must have attributes that allow placement of the event memory in context with the other memories that are already stored in the brain. What are these attributes? There are many on them, including the sensory sources of the event, emotional background etc. The attribute we focus on now is the timing of the event.

The events sequence, what was before and what was after, and the time intervals between the events, are probably the most important events attributes. When I trained my dog to sit, the events were, (A) command "sit", (B) pressing on the dog's back and (C) a sausage. It is a common knowledge that the training will succeed only if the events occur in the sequence, A-B-C, and only if the time-lag between A, B and C is short enough (the shorter the lag the more effective the training). This means that at least for a while the stored in the brain events, A, B and C, have attributes, which we called the time-tags, that allow knowledge of the events sequence and the time intervals between them.

The events train sequence is crucial not only for the Pavlovian training but practically in any experience. When we read the sentence, "snakes eat mice", the resulting information, about who eats and who is eaten, can be obtained only if all three words are analyzed *simultaneously*. Yet the sequence of the words is critical; change of the sequence may lead to a wrong information that mice eat snakes, or to a bad advice to eat snakes and mice. This means that during the *simultaneous* analyses of the words they had some tags that indicated their sequence. But after we analyzed the words in the order they were read and the correct information is stored in the brain, in some hierarchy about animals, food, eating and fairytales from childhood, the words tags become irrelevant; and later we won't even know if we learnt this edifying information via that specific sequence of words, or from seeing a picture or a documentary.

We hypothesize that when a new information, an event or events chain, arrives into the brain, it may reverberate and create some memories in many parts of the brain (as it is discovered by the biologists), but it does not mean that the *same* memories are *duplicated* and stored in parallel in several places, or that the memories are stored initially in the "fast" store and then are copied into (or "train") the other, "slow" store. Instead, we suggest that different storages are used to store different aspects (attributes) of the memories about the same events; the memories that are manifold by their nature, because events are multifaceted. In particular, we hypothesize that the events time-tags are stored in the hippocampus and are used to establish relations between the newly acquired chain of events and pre-existing memories stored elsewhere (neocortex and perhaps some other locations). When the relations are established the time-tags become an obsolete part of the event memory; a burden that hampers the usage of the event when it is recalled via the established relations. If the *absolute* time-tags are used, then, if the time-tag was not removed during NREM sleep when SIT was reset, then the "pre-sleep" time-tag becomes not just obsolete but harmful, because it may lead to a false temporal relation between the new events and the pre-sleep ones. This is why special efforts are applied, in the form of the high amplitude slow waves and other high energy excitations, to remove the time-tags during NREM sleep.

From this perspective, the statement quoted above, "Many aspects of events experienced during waking represent unique and irrelevant information that does not need to be stored long term", is both unspecific and inaccurate. Unspecific, comparing to the specific unique information that we point to, the time-tags. Inaccurate, as at least some of the information does not need to be stored long term not because it is irrelevant *originally*. It might be very important initially to determine the context for the long term storage. Only after the context is found the information is not needed anymore and becomes irrelevant.

In general, the proposed hypothesis provides the same or better explanation of the experimental data than the "two-stage model of model of memory consolidation". Our hypothesis is by far less speculative than the "two-stage model" as it does not imply the existence of some "training" mechanism by which hippocampus trains neocortex. Such a "training mechanism" by itself would be complex and by far not trivial. Instead, we suggest that experimentally observed re-activations are not the "training sessions" but just energy excitations that are needed to expedite destruction (washing off) of the time-tags. The observed diversity of the excitations (spindles etc.) reflects the complexity of the whole brain system and the fact that the temporal attributes (the time-tags) may be attached by different mechanisms to different types of events and other event attributes, e.g. depending on the involved sensory system, and therefore different activation may be needed to make them unstable.[14]

There are variations of the two-stage model or other theories that suggest that "role of sleep is to downscale synaptic strength to a baseline level", or that "information acquired during the waking period is processed first in the early sleep stages, NREM and particularly SWS. Subsequent processing occurs in the later sleep stage, REM, and information eventually emerges in a new form upon awakening". Or, "specifically that, during SWS, slow oscillations, spindles, SWR, and low cholinergic activity all coordinate to promote the reactivation and redistribution of hippocampal-dependent memories to the neocortex, thereby instantiating system consolidation. Subsequently, during REM sleep, high cholinergic and theta activity promote synaptic consolidation of the newly redistributed representations in the neocortex" [29]. However, none of the theories can answer the question why the same functions cannot be performed when the brain is "on-line". Amazingly, most of the stories, both popular and scientific, about sleep imply that the state of sleep is more mysterious than the wakefulness. This, obviously, comes from the naïve fallacy that we know what we do when we are awake. How much do we know about what the brain does when we are awake?

---

[14] Even a simple washing machine goes through a number of the washing cycles, besides the fact that often we have to do special treatment of the dirtiest spots before putting the cloth into the washing machine.

To summarize, the main hypothesis of the suggested theory, about the absolute time-tags – Snippet Inner Time, is derived from the considerations of some specific crucial for the *wakefulness* activities. This hypothesis provides simpler and less speculative explanations of the sleep related experimental results than other theories. Also, so far none of the other theories about sleep function can clearly answer the question, "Why must the brain be taken off-line to accomplish that function?" [1]. This question is directly addressed by our hypothesis that suggests that periodical "taking the brain off-line" occurs because wakefulness is not possible without the specified activities that require *monotonic* change of the inner time; the inner time cannot be monotonic *infinitely* without periods of resetting; and thus when the inner time is being reset wakefulness is turned off.

Another main hypothesis is about the purpose of the REM-NREM sleep cycles. Though we have illustrated possible links between these two hypotheses in their essence they are independent and can be considered standalone.

There are other pieces added to the theory: yawning hypothesis, TBO implementation of SIT hypothesis, yawning mechanism and time-tags location in the hippocampus hypotheses, discussed interpretations of the experimental results etc. They are "extras" that can be skipped, replaced or removed, without changing the theory core. More "extras" can be envisioned; e.g. it seems that the real biological age, how much we lived, in the sense of activity – acquired experience and accomplishments – should be measured not in a calendar units (years), but in the variation[15] of SIT. This might explain why in childhood a calendar year is next to infinity, while after 70 it passes in a glimpse.

## *Author's background and Acknowledgments*

My background is in mathematics, Mechanics of Deformable Solids and software development for Finite Elements Analyses (e.g. [30-34]). Only once, for a short while, I was involved at a professional level in a biology related research project, and that was in Quantitative Epidemiology almost 50 years ago [35]. Though the ideas about the inner time vs sleeping were simmering in my brain for over 30 years (since the amazing time slowing down observations during my "car-crash experiment" in 1987 [Appendix 1]), it was not until reaching the retirement age of 70 that I had time to really pursue them. Inevitably my knowledge of biology is that of an amateur, and my literature research was mostly limited to what is available online, and for this I apologize.

I am grateful to Marat Rvachev [36, 37], Nick Perunov [38] and Timur Rvachov [39] for the discussions of the ideas. I am also deeply grateful to Joyce Gladwell [40] for encouragement and thorough English language editing of much of the paper.

Ontario, Canada, 2020

[15] Here "variation of function" refers to mathematical terminology.

## *References*

1. The Function of Sleep, H.C.Heller, Stanford University, Stanford, CA, USA https://www.sciencedirect.com/science/article/pii/B9780123786104000838?dgcid
2. How does the brain solve visual object recognition? James J. DiCarlo, Davide Zoccolan, and Nicole C. Rust. Neuron, 73(3):415–434, February 9 2012. https://www.cell.com/neuron/fulltext/S0896-6273(12)00092-X
3. On a Time Base of Sounds Recognition by Animals, Michael Rvachov, Appendix 3 below.
4. Procedural Learning: Classical Conditioning, R.E. Clark, R.F. Thompson, in Encyclopedia of Neuroscience, 2009, https://www.sciencedirect.com/topics/psychology/trace-conditioning
5. The Glymphatic System: A Beginner's Guide, Nadia Aalling Jessen, Anne Sofie Finmann Munk, Iben Lundgaard, Maiken Nedergaard, https://link.springer.com/article/10.1007/s11064-015-1581-6
6. Vladyslav V.Vyazovskiy, Umberto Olcese, Yaniv M.Lazimy, Ugo Faraguna, Steve K.Esser, Justin C.Williams, ChiaraCirelli, GiulioTononi, 2009. Cortical Firing and Sleep Homeostasis. Neuron, Vol. 63, Issue 6, pp. 865-878. https://doi.org/10.1016/j.neuron.2009.08.024
7. Circadian variations of adenosine and of its metabolism. Could adenosine be a molecular oscillator for circadian rhythms?" Victoria Chagoya de Sanchez, https://www.nrcresearchpress.com/doi/pdf/10.1139/y95-044
8. Recovery sleep after extended wakefulness restores elevated $A_1$ adenosine receptor availability in the human brain, David Elmenhorst, Eva-Maria Elmenhorst at al., https://www.pnas.org/content/114/16/4243
9. Sleep Deprivation Increases A1 Adenosine Receptor Binding in the Human Brain: A Positron Emission Tomography Study, David Elmenhorst, Philipp T. Meyer at al., https://pdfs.semanticscholar.org/4643/8913aa9fc74fd986a8a3c6cfdf102102160a.pdf
10. Neuropharmacology of Sleep and Wakefulness, Christopher J. Watson, Ph.D., Helen A. Baghdoyan, Ph.D., and Ralph Lydic, Ph.D. https://www.ncbi.nlm.nih.gov/pmc/articles/PMC3026477/
11. Does Time Really Slow Down during a Frightening Event? Chess Stetson, Matthew P. Fiesta, David M. Eagleman https://journals.plos.org/plosone/article?id=10.1371/journal.pone.0001295
12. The two-process model of sleep regulation: a reappraisal, Alexander A. Borbély, Serge Daan, Anna Wirz-Justice ,Tom Deboer, https://onlinelibrary.wiley.com/doi/full/10.1111/jsr.12371
13. A yawning gap, Jim Horn, https://thepsychologist.bps.org.uk/yawning-gap
14. Interplay between Yawning and Vigilance: A Review of the Experimental Evidence, Guggisberg A.G., Mathis J., Hess C.W., https://www.karger.com/Article/FullText/307079#ref21
15. EEG correlats of yawning during sleep onset, John V. Laing, Robert D. Ogilvie, http://baillement.com/eeg.html
16. https://healthjade.net/how-to-sleep-better/
17. http://www.dreamgate.com/dream/library/idx_science_rem.htm
18. http://www.daviddarling.info/images3/loading_cannon.jpg
https://ae01.alicdn.com/kf/HTB1CWgfapTM8KJjSZFlq6yO8FXaq/square-cup-type-mini-Plastic-Washboard-Washing-Board-Shirts-Cleaning-Laundry-For-Kid-Clothes-washing-board.jpg
http://www.rnovelaromantica.com/images/varios/fotosarticulos/lavado13.jpg

19. Forward and reverse hippocampal place-cell sequences during ripples, Kamran Diba, György Buzsáki. Nat Neurosci 10, 1241–1242 (2007). https://doi.org/10.1038/nn1961

20. The inner experience of time. Phil. Trans. R. Soc. B 2009 364, 1955- 1967,doi: 10.1098/rstb.2009.0003, Marc Wittmann, http://rstb.royalsocietypublishing.org/content/364/1525/1955.full.pdf

21. Brain & Cog Sciences and McGovern Institute for Brain Research, MIT. http://nancysbraintalks.mit.edu/, https://www.youtube.com/watch?v=1dkhc8-xuHw&feature=youtu.be

22. Human relational memory requires time and sleep, Jeffrey M. Ellenbogen, Peter T. Hu, Jessica D. Payne, Debra Titone, Matthew P. Walker. PNAS May 1, 2007 104 (18) 7723-7728. https://doi.org/10.1073/pnas.0700094104

23. Sleep-dependent learning and motor-skill complexity, Kenichi Kuriyama, Robert Stickgold, Matthew P. Walker. Learn. Mem. 2004. 11: 705-713. http://learnmem.cshlp.org/content/11/6/705.short

24. Are Spatial Memories Strengthened in the Human Hippocampus during Slow Wave Sleep? Philippe Peigneux, Steven Laureys, Sonia Fuchs, Fabienne Collette, Fabien Perrin, Jean Reggers, Christophe Phillips, Christian Degueldre, Guy Del Fiore, Joe Aerts, Andre Luxen, Pierre Maquet. Neuron. 2004 Oct 28;44(3):535-45. DOI: 10.1016/j.neuron.2004.10.007

25. Labile or stable: opposing consequences for memory when reactivated during waking and sleep, Susanne Diekelmann, Christian Büchel, Jan Born, Björn Rasch. *Nat Neurosci* **14,** 381–386 (2011). https://doi.org/10.1038/nn.2744

26. Odor Cues During Slow-Wave Sleep Prompt Declarative Memory Consolidation, Björn Rasch, Christian Büchel, Steffen Gais, Jan Born. *Science* 09 Mar 2007: Vol. 315, Issue 5817, pp. 1426-1429. https://science.sciencemag.org/content/315/5817/1426

27. The memory function of sleep, Susanne Diekelmann and Jan Born. *Nature Reviews Neuroscience*, Vol. 11, pp.114–126(2010). https://doi.org/10.1038/nrn2762

28. Sleep-dependent directional coupling between human neocortex and hippocampus, Tobias Wagner, Nikolai Axmacher, Klaus Lehnertz, Christian E.Elger, Jürgen Fell. Cortex, Vol. 46, Issue 2, Feb. 2010, pp. 256-263. https://doi.org/10.1016/j.cortex.2009.05.012

29. The Restless Engram: Consolidations Never End, Yadin Dudai . Annual Review of Neuroscience. Vol. 35:227-247. https://doi.org/10.1146/annurev-neuro-062111-150500

30. Rvachev M. A. 1973, On the solution of some Monge-type problems. Dokl. Akad. Nauk SSSR, 1974, Volume 219, Number 1, pp 46–48, http://www.mathnet.ru/php/archive.phtml?wshow=paper&jrnid=dan&paperid=38640&option_lang=eng

31. Rvachev M. A. 1977, A problem of Ulam type. Functional Analysis and Its Applications, April 1977, Volume 11, Issue 2, pp 128–135, https://link.springer.com/article/10.1007%2FBF01081891

32. Rvachev M. A. 1990, On statically possible fields in a simply connected volume. Journal of Applied Mathematics and Mechanics, Vol 54, Issue 1, 1990, pp 143-145, https://doi.org/10.1016/0021-8928(90)90100-O

33. Pokras V.D., Rvachev M.A. 1996, Application of the R-functions method to visioplastic analysis in metal forming. Journal of Materials Processing Technology, Vol 60, Issues 1–4, pp 493-500, https://doi.org/10.1016/0924-0136(96)02376-X

34. Wang J., Rvachov M.A., Hou T. 1999, 2D Finite Element Simulation of Sheet Metal Forming Processes. SAE Technical Papers, March 1999, https://doi.org/10.4271/1999-01-1004

35. Computer Modelling of Influenza Epidemics for the Whole Country (USSR), O. V. Baroyan, L. A. Rvachev, U. V. Basilevsky, V. V. Ermakov, K. D. Frank, M. A. Rvachev and V. A. Shashkov, Advances in Applied Probability, Vol. 3, No. 2 (Autumn, 1971), pp. 224-226 https://www.jstor.org/stable/1426167?seq=1#page_scan_tab_contents

36. On axoplasmic pressure waves and their possible role in nerve impulse propagation, Marat Rvachev, Biophys. Rev. Lett. 05, 73–88 (2010), https://doi.org/10.1142/S1793048010001147

37. Neuron as a reward-modulated combinatorial switch and a model of learning behavior, Marat Rvachev, Neural Networks, 46, 62-74 (2013), https://doi.org/10.1016/j.neunet.2013.04.010

38. Statistical Physics of Adaptation, Nikolay Perunov, Robert A. Marsland, and Jeremy L. England, Phys. Rev. X **6**, https://journals.aps.org/prx/abstract/10.1103/PhysRevX.6.021036

39. Rvachov T.M., Son H., Sommer A.T., Ebadi S, Park J.J., Zwierlein M.W., Ketterle W, Jamison A. O 2017, Long-Lived Ultracold Molecules with Electric and Magnetic Dipole Moments. Phys. Rev. Lett. **119**, 143001, https://link.aps.org/doi/10.1103/PhysRevLett.119.143001

40. Brown Face Big Master, Joyce Gladwell, https://www.amazon.com/gp/product/0333974301/ref=dbs_a_def_rwt_hsch_vapi_taft_p1_i0

41. Wackermann J., Ehm W. 2006, The dual klepsydra model of internal time representation and time reproduction. Journal of Theoretical Biology, Vol. 239, Issue 4, pp. 482-493, https://doi.org/10.1016/j.jtbi.2005.08.024

42. Muheim C., Brown S.A. 2013. Adenosine and Other Purinergic Products in Circadian Timing. DOI:10.5167/uzh-74287 https://www.pharma.uzh.ch/dam/jcr:00000000-65ce-162b-ffff-ffff81b32991/MuheimBrown.pdf

43. Dorato M., Wittmann M. 2019. The phenomenology and cognitive neuroscience of experienced temporality. Phenomenology and the Cognitive Sciences https://doi.org/10.1007/s11097-019-09651-4

44. Marc Wittmann, Virginie van Wassenhove, A. D. (Bud) Craig and Martin P. Paulus. 2010. The neural substrates of subjective time dilation. Front. Hum. Neurosci. https://doi.org/10.3389/neuro.09.002.2010

45. Marc Wittmann, Virginie van Wassenhove, 2017. Why Time Slows Down during an Accident. Neuroscience. Front. Young Minds. 5:32. https://doi:10.3389/frym.2017.00032 https://kids.frontiersin.org/article/10.3389/frym.2017.00032

## *Appendix 1. Does Time Really Slow Down during a Frightening Event?*

The research [1] named as above tries to prove that it does not. Let us show, based both on subjective and objective observations, that the research does not prove anything. The approach was interesting and daring, as the participants had to free fall from of a Suspended Catch Air Device (SCAD) tower. Here is a short extract from the research [1]. *"Observers commonly report that time seems to have moved in slow motion during a life-threatening event... Using a hand-held device to measure speed of visual perception, participants experienced free fall for 31 m before landing safely in a net. We found no evidence of increased temporal resolution, in apparent conflict with the fact that participants retrospectively estimated their own fall to last 36% longer than others' falls...Our findings suggest that time-slowing is a function of recollection, not perception… We speculate that the… emotional memory may lead to dilated duration judgments retrospectively"*.

Let me start with my subjective objections to the research conclusion, which originate from the car crash that I was in about 30 years ago. Two circumstances do not fit into the research assumptions. First, there was no a subjective premise, an expectation of the coming accident that could create some special emotional background of the event. I did not see it coming and did not know that it was a car crash until many hours later, when the car was raised and I saw that the front axle beam was warped into almost a pretzel. Only then I realized that it was really a life-threatening crash and that I was saved only because I was buckled up.

The details of the crash may be interested to many readers, as I would guess most would never imagine that this was possible. It was in the U.S.S.R in 1987. A construction company was adding one more lane to the road. When they ran out of the money for the project they dropped the new lane construction halfway, until the funds to extend the lane farther would be available. They removed their equipment, all the temporary fencing that was protecting the worksite while they were working, and went away. There was no a smooth merge of the new lane ending with the old road. Not a single sign was left to indicate that the lane ends. There was no a lane marking at all.

When I was driving, at night, as I thought in the middle of the empty lane, the front axle beam hit a practically absolutely rigid obstacle, a row of the curb stones edging the old road. My car was pinned by the curb stones like a fly is pinned for the insect collection. After the crash, from outside the car looked not much damaged (until the car was raised and we saw the pretzel axle), though the engine compartment shifted a little, which lead to the windshield fracturing. Inside, the rear seat back was torn off its constraints and became a ramp over which the stuff from the luggage compartment was unloaded onto me.

I called the police. When the officer arrived he began to lament[16] that he had tired documenting the accidents on this very spot. Only during his shifts there were more than five of them, including a motorcyclist who broke his spine.

Can somebody imagine anything like this in a safe country? The construction company would be out of business and somebody would have been criminally charged, thanks to the lawyers.

Anyway, back to the research discussion. My second subjective experience that does not fit the research conclusion is the suggestion that the time-slowing is a function of recollection. By the time of the accident I had seen many a specimen destruction in a test machine, which was a part of my professional experience, and had some definite ideas about the fracture propagation speed. So, when during the "my car crash experiment" I saw a fracture in the windshield that was propagating and branching so slowly that I could see all the details, I was amazed *right away*. Another strange thing that was happening was a heavy metal thermos filled with 5 litters of tea that arrived from the luggage compartment and was hovering near my head as a toy balloon, though at the time most of my thinking was about the fracture as I was not interested in the balloons.

Now, the *objective* objection to the research conclusion. The speed of visual perception was measured by a hand-held device with a flickering dial. The flickering frequency was adjusted so that the experiment participant could not see the flickering, but a tiny drop of the frequency would make the flickering visible. The assumption was that if *"time as a whole runs in slow motion during frightening events"* then during the fall from the SCAD tower the participant would see the flickering though its frequency was not changed. As it is correctly noted in the research discussion, *"A critical point for the logic of this study is that flicker fusion frequencies are not limited by the retina"*, and they come to the conclusion that they are not, *"since retinal ganglion cells have extremely high temporal resolution"*.

This last conclusion is an erroneous understanding of what kind of temporal resolution is required to see *flickering*. To see flickering, the eye retina/brain system should react and we should see the light when it appeared, and the system should recuperate and we should stop seeing the light when it is gone. The recuperation time is many times longer than the excitation time, as almost all of us observed on many occasions. Imagine that somebody is waving a flashlight at night. What do we see? If waving is slow, we see the light as one spot, because it is in one spot. But as waving gets faster, the spot becomes a ribbon, and it could be a pretty long ribbon comparing to the size of the flashlight, Fig.1.

[16] This is a delicate description of what the officer was saying.

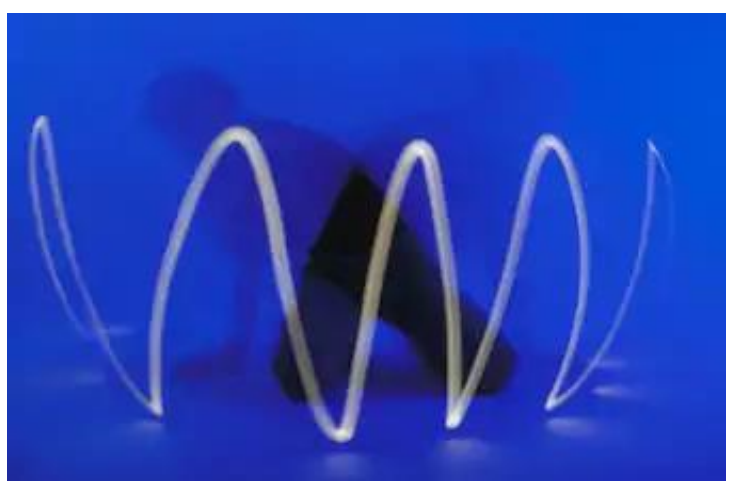

Fig.1. A moving flashlight may look like a ribbon, https://www.shutterstock.com/image-photo/man-waving-flashlight-blue-background-387206. Of course, on the photo it is because of the photo-exposure duration. But we can see a similar picture "in vivo", without a camera[17].

The sensitivity to flickering is limited by the recuperation time. The fact that we can see the head of the ribbon far apart from its tail shows that our brain can work much faster than retina/brain system can recuperate. Flickering could not detect a change in our "brain speed", even if it occurred during the free fall.

Then the only conclusion we can derive from that research is that the participants, subjectively, observed that during the free fall time slowed down 36%. Since there were several participates who felt this way, maybe there was some objective reality behind these observations.

## ***References***

1. Does Time Really Slow Down during a Frightening Event? Chess Stetson, Matthew P. Fiesta, David M. Eagleman https://journals.plos.org/plosone/article?id=10.1371/journal.pone.0001295

[17] When I was a teenager, at a bonfire at night, we were pulling out sticks with the smoldering ends out of the fire and competing whose waving ribbon was longer.

## *Appendix 2. Is Cortical Firing Variable Frequency an Inner Time?*

***Abstract***

It was found by Vladyslav V. Vyazovskiy and his team in 2009 that the firing frequency of cortical neurons increases monotonically during wakefulness. However the significance of this discovery is not yet known: is the frequency change a self-important phenomenon or an insignificant side effect of sleep-wake homeostasis? We provide a hypothesis that the observed monotonic frequency change may be a part or an echo of a resonance-based inner clock mechanism. If the hypothesis is correct then it explains the necessity of sleeping.

***Introduction***

There is an enormous amount of research about the inner time, but "There is yet no conclusive answer as to what kind of time keeping mechanism is implemented in the brain" [1]. Our goal is to show that it is plausible that the time keeping mechanism was already observed and that "inner clock ticking" was measured by Vladyslav V. Vyazovskiy and his team in 2009 [2], though the putative significance of their observations is discovered here for the first time.

The cortical neurons activity was studied in [2] on rats, but it is likely a similar pattern exists in many other species, including humans. We narrow the discussion to humans, because there is no need to talk more generally to convey the ideas below.

Normal awake functions of a human include hearing, seeing, moving and keeping an upright standing position.

To recognize sounds, the same sound sequence repeated several times should unfold in the brain the same way each time, which means that the internal "sound recording mechanism" should always run with the same speed vs the real[18] time. This means that hearing requires a proper inner time to control internal unfolding of sounds.

To visually recognize the velocity of moving objects and react properly, the internal velocity measuring and reaction mechanisms should be controlled by the inner time that properly reflects the real time increments.

In general, any learned muscle activity – including the ability to keep an upright standing position – must also be controlled by the inner time that converts real time intervals into inner time intervals properly, because the velocity increment of an object is proportional to the applied force and its duration (second Newton law), and body parts are mechanical objects that obey the Newtonian mechanics.

The listed activities require an internal mechanism that measures (in some internal "units") the time interval between recent events (stimuli) and "now", the present moment. This topic of neural processing of time has been widely discussed, but a satisfactory timing model (mechanism) was not yet found [3]. Next we provide a novel specific hypothesis about such a mechanism and then compare its functioning with experimental observations.

---

[18] as opposed to inner

***Time-Bearing Oscillator model of timing***

Let us hypothesize that there is a *variable frequency* oscillator contour in the brain, the frequency of which changes over the real time monotonically[19] and incrementally. Let us call it a "time-bearing oscillator", TBO. Suppose that when each input event (stimulus) arrives – i.e. registered by the brain – a "tuning fork" resonator is added to the event memory as the event time-tag. The tuning fork, TF, has a *fixed natural frequency* that is set to be equal to the TBO frequency at the moment of the event arrival. Thus the natural frequencies of the recorded (memorized) TFs will differ from the "now" frequency of the time-bearing oscillator, TBO; the earlier the TF was set the larger the difference. In addition, let us assume that the oscillation of the TBO is imposed to excite the TF and creates a forced oscillation in the TF. The amplitude of the forced oscillation, compared to the excitation amplitude (transmissibility), can serve as the internal measure of the time difference between "now" and the time of the event arrival – the larger the amplitude the more recent the occurrence of the event.

We call the period, $[t_0, t_1]$, during which the TBO frequency is monotonic and incremental, "TBO temporal range", and TBO frequency change, $[\omega_0, \omega_1]$, during this period – "TBO frequency range".

To assess the sensitivity of such a measurement let us assume that the TF is a harmonic oscillator; TF has natural frequency $\omega_R$, ($R$ stands for "resonator"), and oscillations of TBO are given by

$$x_{TBO} = F\sin(\omega\, t), \text{ where } \omega = \omega(t),\ t_0 < t < t_1\ ,$$

while oscillations of TF are forced oscillations defined by the equation

$$\frac{d^2 x_{TF}}{dt^2} + 2\zeta\omega_R\frac{dx_{TF}}{dt} + \omega_R^2 x_{TF} = x_{TBO}\ , \qquad (1)$$

where ζ is TF's damping [4].

If $\omega(t) = \omega = const$ then (1) has a closed-form solution

$$x_{TF} = \frac{F}{K}\sin(\omega t + \varphi)\ , \qquad (2)$$

where $K = \sqrt{(2\omega_R\omega\zeta)^2 + (\omega_R^2 - \omega^2)^2}$ and $\varphi = \arctan\left(\frac{2\omega_0\omega\zeta}{\omega^2 - \omega_R^2}\right) + n\pi$.

We'll use the solution (2) to approximately evaluate $x_{TF}$ when $\omega(t)$ is slowly varying; and afterwards discuss the shortcomings of this approach.

Suppose that the TBO temporal range is the period of wakefulness, and $\omega(t)$ is linear, $\omega(t) = \omega_0 + ct,\ c = const$. Let us explore the TF amplitude change as a function of the time interval, $\Delta t$, from the event arrival to "now", assuming that $|c\Delta t| \ll \omega(t)$. Within the assumption $\omega(t)$ is almost constant, $\omega(t) \approx const$, and $x_{TF}$ is approximately defined by (2). Examples of the TF amplitude are shown on Figure 1.

The TBO frequency range at the bottom of Figure 1 is one of the possible cortical neuron firing variable frequency ranges in rats, derived from [2]; while the range on the top was selected, because human EEG appears to be the most dependent on the prolonged wakefulness duration in this range (e.g. [5]) and we did not have any better hint of how to select a range that might be applicable to a human.

---

[19] "Monotonically" is used here in a narrow mathematical meaning of a monotonic function.

The graphs' shape critically depends on the damping parameter $\zeta$. In Figure 1 we selected $\zeta = 10^{-6}$, because close to this $\zeta$ value, the 10ms time intervals produce clearly detectable differences in the TF amplitude within 500 ms from "now", while farther from "now" the sensitivity of the TF resonator to the time interval duration sharply drops. Qualitatively this is consistent with trace conditioning and other observations of the sensorimotor temporal discrimination sensitivity in humans [3].

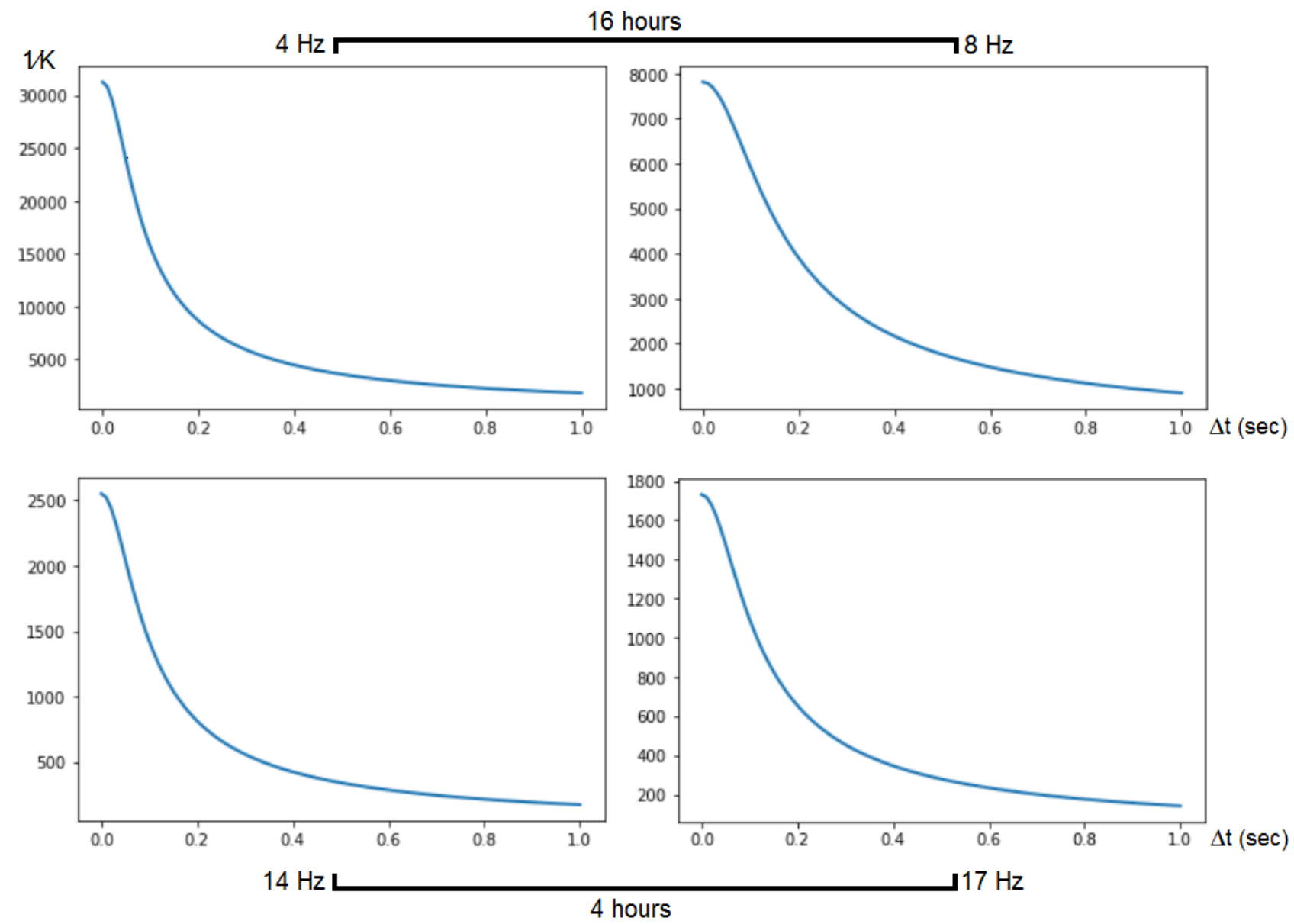

Figure 1. Transmissibility, 1/K, of the TBO oscillations onto the TF, as a function of the time interval, $\Delta t$, between the TF-event registration and "now", in the time range up to 1 second from TF-event registration. After TF-event registration, the TBO frequency continues to increase linearly, $\omega(t) = \omega_0 + ct$, at a rate of c = 0.25 Hz / Hr (top), and c = 0.75 Hz / Hr (bottom). As the TBO frequency increases, it is further off-resonance from the TF and the decaying amplitude[20] of the driven TF oscillation can be used as a measure of time differences up to approximately 500 milliseconds, after which the TF response is small and little-changed with elapsed time since event registration. Within 500 milliseconds, 10 ms time increments produce clearly distinguishable TF excitations. The TF damping coefficient is $\zeta = 10^{-6}$.
Left – at the beginning, right – at the end of the wakefulness period.
Top: TBO changes from 4 Hz to 8 Hz over 16 hours.
Bottom: TBO changes from 14 Hz to 17 Hz over 4 hours.

This example illustrates that the TF-time-tag forced oscillation amplitude might serve as a very sensitive and accurate measure of the time intervals from "now" to recent events, within 500 ms snippets from "now", at any moment of the whole period of wakefulness. A very detailed analysis of the Figure 1 graphs is not useful, however, because brain oscillations can hardly be exactly modeled by equation (1), the parameter selection in these examples is intuitive but somewhat arbitrary, and any process in a living being is hardly exactly linear over a prolonged time. Still two observations may be of note: First, to detect the differences between 10 ms events, the TF resonator must be quite efficient (low damping[21]), and in this case the forced oscillation amplitude might be several orders larger than the excitation amplitude. This means that the TF echoes of the TBO might be by orders "louder" than TBO itself, which makes an experimental TBO discovery quite challenging. Second, if the damping is constant then towards the end of wakefulness the sensitivity of the temporal detection drops. Overall this is in compliance with the observation that at the end of a prolong wakefulness we are not as mentally sharp as at the beginning.

[20] If the decay can be observed during the observation interval (see the discussion below).

The illustrated TF sensitivity is not intuitive, because in the examples in Figure 1 the TBO frequency changes extremely slowly, $7 * 10^{-5}\ Hz/sec$ on the top and $2 * 10^{-4}\ Hz/sec$ on the bottom, which means that the TF resonator must capture the frequency changes of the order of $10^{-7}\ Hz$. However, we have shown above that this does occur, near the TF natural frequency, if the TF is a high quality resonator, with the damping around $10^{-6}$.

Can the high quality TF resonator exist? The resonators with damping around $10^{-6}$ are known to exist naturally; e.g. quartz crystal oscillator has damping in this range (though it has a completely different natural frequency). Nothing is known about the nature of the putative TF, besides the fact that the brain is a very efficient device, and that nature often re-uses the same ideas in different areas and on different scales. *A priori*, we cannot dismiss the possibility of the existence of a high quality TF oscillator.

Then there is a question if the TF amplitude decay can be observed during the observation time (10 ms) that is much shorter than the $\omega_R$ wave length (~100 ms). The question has two sides: First, if the TF oscillation maximum will fall into the observation interval and thus can be observed and, second, if the decay of the driven oscillation will not be masked by the decay of the natural TF oscillation, which will occur at the rate $e^{-\omega_R \zeta t}$ that is much slower than the decays that are shown on Figure 1. The first side might be addressed if the TF is implemented not as a single resonator but as a "population of resonators" – an incoherent ensemble of the resonators of the frequency $\omega_R$ that are being exited by $\omega(t)$ with the scattered delays. Then the decaying *ensemble* excitation might be used as a measure of time intervals, even if they are shorter than $\omega_R$ wave length, depending on the number of the resonators in the ensemble and how densely each resonator maximums populate the oscillation period. To address the second side, let us note that even if a TF resonator exists, when it is added to the event memory as the event time-tag, the addition itself will change the TF properties, because association with the event memory would require the TF modification and including the TF into a number of other oscillation loops. Thus modelling the TF as a harmonic resonator controlled by the equation (1) is flawed from the beginning. The provided discussion might be qualitatively but not quantitatively interesting.

Still let us add two more observations within this TF model. While the TBO frequency changes monotonically and incrementally, the frequency *derivative* might change *not* monotonically or incrementally, and be quite "jumpy", depending on circumstances, e.g. if are we focused or relaxed, in a car crash or close to napping. The dependency of the TF time interval discrimination on $\dot{\omega}$ is illustrated on Figure 2, left. The TF interval discrimination is highly sensitive to the derivative of the TBO frequency; a higher derivative results in a finer interval discrimination. A significant increase of $\dot{\omega}$ will result in the perception of the world in a "slow motion". E.g. on the Figure 2, left, $\dot{\omega} = 2 * 10^{-4} Hz/sec$ and a ~200 ms real time interval is measured as 750 "internal units" (the ratio of the TF amplitude to the TBO amplitude). When the derivative is 4 times larger, the same 750 units will correspond to a ~50 ms real time interval, which means that the world will be perceived as moving ~4 times slower. Such "slow motion" time dilation has been experimentally studied in [6], among many other observations. The decrease of the TBO frequency derivative, obviously, produces an opposite effect. A key observation here is that the *derivative* of TBO frequency is responsible for inner time measurement resolution. The role of the TBO frequency *per se* is that it sets the limits, $[\omega_0, \omega_1]$, within which the TBO clock can operate. When we are active the demand for inner temporal resolution is high, which requires a higher $\dot{\omega}$ and leads to a faster reaching of the TBO, $\omega(t)$, frequency limit and the necessity to reset the TBO. We hypothesize that the TBO frequency is reset during sleep, while sensorimotor functions are inactive and there is no need for a forward-moving TBO frequency. If the TBO clock is the wakefulness timing mechanism that would mean that higher wakeful activity—and thus higher required temporal resolution, — results in a sooner necessity to sleep. Inversely, if the activity is low the demand to sleep might not come from the TBO clock at all, while other factors, like the circadian rhythm, might weigh in, and we may need special means (sleeping pills etc.) to fall asleep.

[21] A resonator efficiency is also often characterized by the Q factor (quality factor), an inversion of the damping.

Note that if one draws a graph of the TBO frequency these variations of the frequency derivative will be hardly noticeable, because the derivative (~$10^{-5}$/sec$^2$) is several orders smaller than the (squared) frequency (~100/sec$^2$). For a study of the alertness variation, the graph of the TBO frequency derivative should be considered separately from the graph of the TBO frequency, in its own scale.

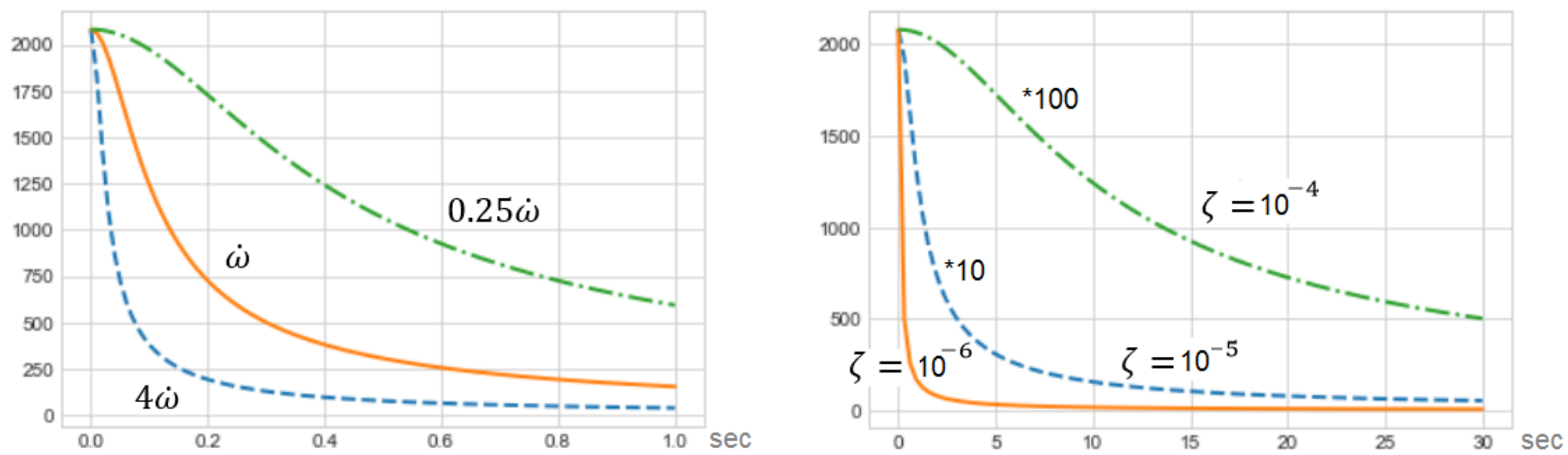

Figure 2. Transmissibility, 1/K, of the TBO oscillations onto the TF as a function of $\Delta t$ (see Figure 1), depending on $\dot{\omega}$ and $\zeta$, in the middle of the TBO 4 hour temporal range (frequency range from 14 Hz to 17 Hz).

Left: 4-fold increase of $\dot{\omega}$ (of the $\omega(t)$ graph slope) from the average, $2 * 10^{-4} Hz/sec$, produces a sharp increase in temporal sensitivity (we see world in a slow motion) while a 4-fold decrease produces an opposite effect.

Right: Increase of the damping moves the sensitivity range from milliseconds ($\zeta = 10^{-6}$) to 10s of seconds ($\zeta = 10^{-4}$). For $\zeta = 10^{-5}$ and $10^{-4}$ the graphs are scaled up 10 and 100 times accordingly.

On Figure 2, right, we show the dependence of the sensitivity on the TF damping. This illustrates that the TBO clock may provide the time interval discrimination in the ranges from milliseconds to tens of seconds, depending of the TF damping.

Thus much of the human temporal dependencies may be provided by the TBO clock, if it exists. Next we refer to the experimental results that can be interpreted either as a hint that the TBO clock exists, or as a direct confirmation, if we assume that nature would not in vain support a monotonic incremental oscillation in the brain.

***Is cortical neuronal firing frequency TBO or a TBO trace?***

Below are several related quotes from [2].

"The changes in neuronal firing patterns and firing rates reported here appear to follow SWA, suggesting that they are homeostatic. To test the homeostatic nature of these effects and experimentally disentangle them from circadian effects, we performed 4 hr of sleep deprivation starting at light onset and measured cortical activity during the subsequent recovery sleep. Sleep deprivation was successful (rats were awake 93% of the time) and, as expected, recovery sleep was associated with elevated SWA compared to the corresponding time of day during baseline." "During sleep deprivation, neuronal firing rates showed a progressive increase up to the third hour and then reached a plateau during the fourth hour. To investigate in more detail which changes in neuronal activity patterns could account for this time course, we quantified the number of long (>50 ms) and short (<20ms) interspike intervals (ISIs) for each individual neuron. Both measures increased progressively, suggesting that while neurons showed increased firing as reflected in the short ISIs (>40% increase), this increase was counteracted at the end of sleep deprivation by an increased number of neuronal silent periods" (Figure 3).

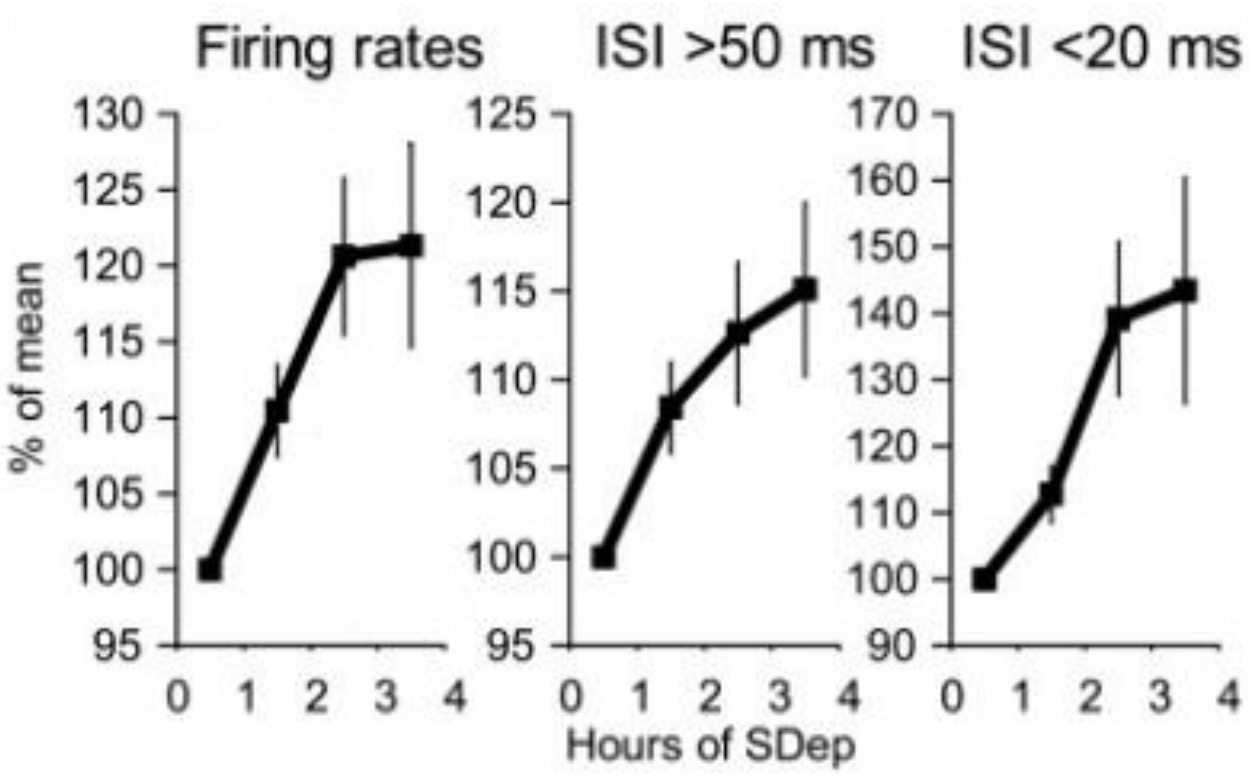

Figure 3, [2]. Effects of Sleep Deprivation on Cortical Firing. Time course of neuronal firing rates, and the number of long (>50 ms) and short (<20 ms) interspike intervals (ISIs) in waking during SDep (50 neurons, n = 5 rats). Mean values ± SEM shown as % of the value during the first hour of SDep.

"We found that the wake-related increase in firing rate was present even after short waking bouts but only if waking was not interrupted by sleep attempts… At the same time, even short (<15 min) "power naps" led to a significant decrease in firing rates, although the effect was more pronounced after longer sleep periods."

"On average, firing rates were ∼15%–20% higher during active relative to quiet waking (15.4 ± 0.9 versus 13.0 ± 0.8 Hz)".

In the research [2] they did not search specifically for inner time carriers. Yet quoted findings suggest that there are oscillatory processes in the brain, frequencies of which change monotonically and incrementally during wakefulness, while during sleep the frequencies are being reset. Each oscillator with such a frequency potentially can serve as the TBO or can be an echo of the TBO, with the wakefulness period being the temporal range.

Of course, by itself the existence of a neuronal oscillator that changes its frequency monotonically during wakefulness does not prove that its frequency at the time of the event arrival is attached to the event memory as the time-tag, in the form of the TF, and that the oscillator is a TBO. However, at least *some* time-tag *must* be attached to the event memory, so that events could be temporally differentiated. The minimum complexity of storing the TF (tuning fork) description as a combination of the brain receptor states is the same as the complexity of storing any description of "now", because the minimum complexity is ultimately defined only by the required temporal discretization. Would it not be likely that evolution has found a way to use the monotonic incremental variable frequency oscillator – which, as we already know from the experiment, exists in the brain – to work as the TBO, because it provides a simple instantaneous tool to temporally differentiate the events, sorting them by the strength of their echoes to "now" (to the TBO of the present moment), emphasizing recent events and attenuating old ones? What could be another purpose of such a variable frequency oscillator?

***Are many TBOs used for timing of various activities?***

Above we considered putative TBO with the temporal range consisting of the whole period of wakefulness.  However if a TBO realization on a neuronal level is possible then one could expect that the same mechanism will be reused for timing of different short-time activities, with the duration of the activity as TBO temporal range. In [7] they studied neural activity in the parietal cortex of monkeys during a time reproduction task and observed monotonic changes of the firing rate during the different steps of the task, e.g. Figure 4. Could the observed monotonic frequency change be a trace of the TBO with the temporal range of 100 – 1100 ms and the frequency range 30 – 80 Hz? Three deviations from the monotonic change, observed on the Figure 4, at the start and the end of the task and 100 ms after "Set", may be the periods before and after the monkeys began working on the timing task, and during the switch from the measurement to the production epoch, when the usage of the timing was changed.

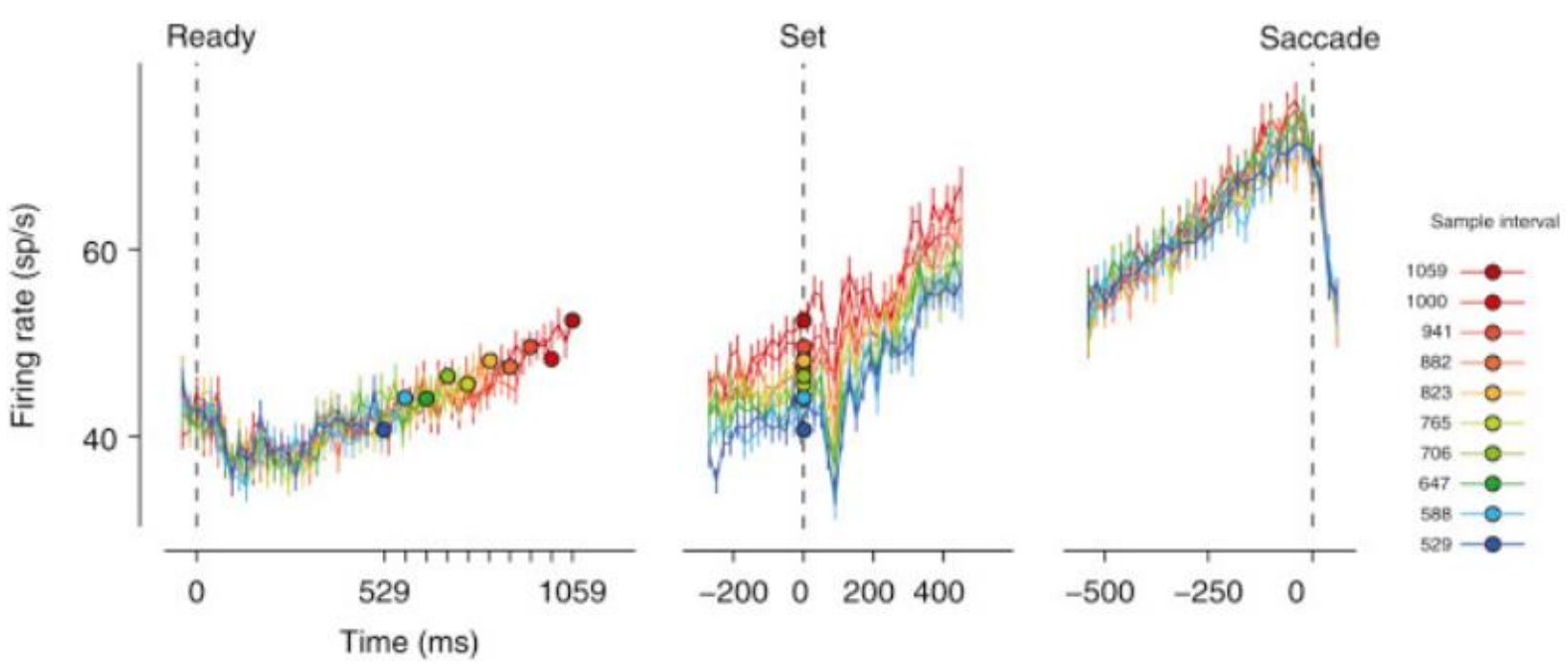

Figure 4 [7]. LIP neurons firing frequency at different steps of the task.

***Summary.***

We have proposed the Time Bearing Oscillator timing model as a mechanism for inner time measurement. We hypothesize that there is a TBO-based internal clock that must run during the period of wakefulness, which enables full-fledged awake functioning, including the ability to recognize sounds, visually recognize movements and keep an upright standing position. The necessity to reset the TBO frequency – to rewind the inner clock – is the reason for the existence of sleep as an inhibited behavioral state. We also hypothesize that many different TBO-type timing mechanisms might be used in different activities.

Comparing to other models of timing mechanisms [3], TBO clock does not require a counter or an accumulator, because it assesses time intervals via resonance, and TBO clock is "natural", because forced oscillations and (near) resonance conditions are pervasive in nature. Oscillations in the brain with a monotonically changing frequency, during the whole period of wakefulness or during specific tasks, have already been discovered experimentally. These oscillations might be a precursor for the existence of a TBO. For now, the TBO clock is the only timing model that has at least a hint of an experimental confirmation of a neuronal level.

***References***

1. Marc Wittmann, Virginie van Wassenhove, A. D. (Bud) Craig and Martin P. Paulus. 2010. The neural substrates of subjective time dilation. Front. Hum. Neurosci. https://doi.org/10.3389/neuro.09.002.2010

2. Vladyslav V.Vyazovskiy, Umberto Olcese, Yaniv M.Lazimy, Ugo Faraguna, Steve K.Esser, Justin C.Williams, ChiaraCirelli, GiulioTononi, 2009. Cortical Firing and Sleep Homeostasis. Neuron, Vol. 63, Issue 6, pp. 865-878. https://doi.org/10.1016/j.neuron.2009.08.024

3. Michael D. Mauk, Dean V. Buonomano, 2004. The neural basis of temporal processing. Annual Rev. Neurosci., 27 (2004), pp. 307-340. https://doi.org/10.1146/annurev.neuro.27.070203.144247

4. Grant R. Fowles, George L. Cassiday, 1986. Analytic Mechanics (5th ed.), Fort Worth: Saunders College Publishing, ISBN 0-03-96746-5, LCCN 93085193

5. Sara Fattinger, Salome Kurth, Maya Ringli, Oskar G. Jenni, Reto Huber, 2017. Theta waves in children's waking electroencephalogram resemble local aspects of sleep during wakefulness. https://www.nature.com/articles/s41598-017-11577-3.pdf

6. Marc Wittmann, Virginie van Wassenhove, A. D. (Bud) Craig, Martin P. Paulus, 2010. The neural substrates of subjective time dilation. Front. Hum. Neurosci. https://doi.org/10.3389/neuro.09.002.2010

7. Mehrdad Jazayeri, Michael N.Shadlen, 2015. A Neural Mechanism for Sensing and Reproducing a Time Interval. Current Biology, Vol. 25, Issue 20, pp. 2599-2609. https://doi.org/10.1016/j.cub.2015.08.038

8. Vladyslav V. Vyazovskiy, Irene Tobler, 2005. Theta activity in the waking EEG is a marker of sleep propensity in the rat. Brain Research 1050, pp. 64 – 71. https://doi.org/10.1016/j.brainres.2005.05.022

## *Appendix 3. On a Time Base of Sounds Recognition by Animals*

### *Abstract*

Sounds are information sequences that cannot exist outside of a time base and therefore cannot be analyzed inside an animal without an accurate internal clock. It is suggested that the evolutionary development of the clock may be linked to the development of the inner ear. The causes of vertigo during rotation are discussed. It is shown that if a continuous inner time exists then sleeping is a "mathematical necessity". The "time slowing down" during an extreme event is discussed. A resonance based mechanism of the time snippets measurements is hypothesized. The concept of the Time Intervals Pinned activities and Snippet Inner Time is introduced. Yawning and sleeping hypothesis are suggested.

### *Introduction*

As it is noted in the fascinating review *The inner experience of time* [1], "the striking diversity of psychological and neurophysiological models of „time perception“ characterizes the debate on how and where in the brain time is processed". Despite the diversity, so far not a single concrete internal process has been identified that cannot function without an inner time, aside from the generic understanding that "the perception of time is an essential and inextricable component of everyday experience". This probably is a major factor why "no conclusive answers to the questions of which neural substrates and what kind of neurophysiological processes could account for the experience of duration have been established", because without a specific understanding where the internal time is used it is not clear where to search for its physiological mechanism.

In this paper, we focus on the observation that hearing cannot function without an accurate inner time. Sounds are information sequences that are distributed in physical (vs *inner*) time – and do not have meaning without a time base that is coherent with physical time, – and therefore cannot be analyzed internally without an inner time that is accurate enough vs the physical time. Hearing could not have been developed evolutionarily without a concomitant inner clock and we discuss the idea that the auditory system includes an inner clock in its mechanism.

### *Hearing inner time*

Fig.1 is a recording of the spoken word "sound" made by a modern computer with the "Audacity" software [2]. The whole word lasts 0.8s.

By replaying different length pieces of the recording we estimated that the length of the replayed piece, to make the word "sound" recognizable, must be at least 0.6s. This means that to be recognized the recording must be analyzed *as a whole*, as a function known in a large enough time interval.

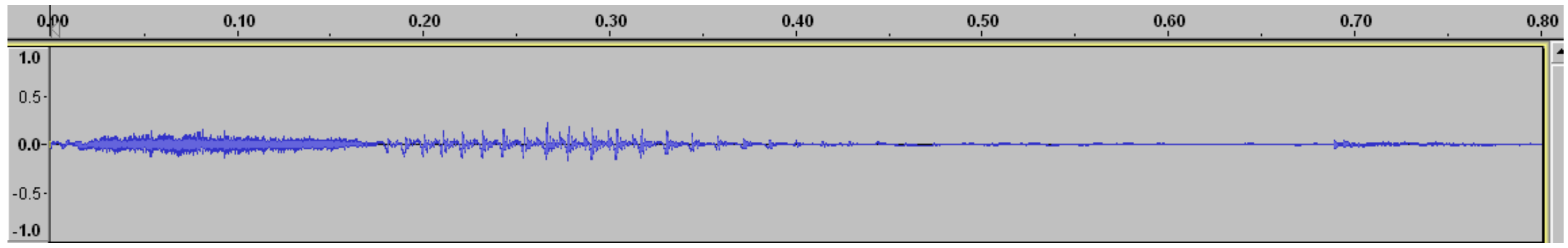

Fig1. Recording of the word "sound". The horizontal axis is (computer) time in seconds (we do not differentiate between physical time and computer time since they are synchronized with much higher precision than relevant to this discussion). The vertical axis is the amplitude of the electrical signal registered by the microphone, normalized by the maximum value set by the software.

The most interesting thing about the recording is that no matter how many times it is replayed to a person the person always hears the same; same word, voice, pitch and tempo. It might appear trivial, but in fact it means that there is an inner time that acts exactly as the real time, at least within short intervals.

Each replay produces an external to the person information sequence (sound waves sequence), a function of physical time. Hearing converts this function into its image in the person's brain. Since the person hears the same each time, the conversion of this external function of time produces the same internal image. The image must be analyzed as a whole to be recognized. The image is a sequence of the distributions of the "instantaneous" sound frequencies and their amplitudes. The sequence itself is a distribution of the input information over the *brain's* time, which is the *inner* time. Since we don't know how may inner clocks and times an animal has, we will refer to the inner time used to analyze hearing as hearing inner time, HIT. So we have a distribution over HIT of the instantaneous amplitude-frequency distributions.

For the image being the same two things must occur. First, each instantaneous amplitude-frequency distribution must be the same for the same input. This occurs because the stratification of the incoming sound by pitch (by frequency) depends on which part of the cochlea responds (resonates and produces electrical signal to the CNS) to the pitch. This remains constant for as long as the electro-mechanical properties of the ear remain the same.

Second, the distribution over HIT of the instantaneous amplitude-frequency distributions must be the same for each replay. For this during the conversion equal length physical time intervals must be converted into equal length HIT intervals. This means that HIT(t) must be a linear function of (physical) time, $HIT(t) = Kt$, maintaining the same coefficient of the proportionality, the HIT slope $K$, during all the replays.

Comparing the conversion of the sounds by hearing to the sound recording by any recording device, a consistent recording is possible only when the recorder timebase speed (e.g., the speed of the movement of the vinyl record groove vs the needle) is constant enough vs physical time. Note that this comparison is not exactly accurate, because the change of the recorder timebase speed during the replay will change both the pitch (sound frequencies) and the tempo (the distribution of the frequencies over the time interval). Contrary to this, if the HIT "timebase speed" (which is the HIT slope) changes it will result only in the change of the tempo without changing the pitch, as the latter is defined by the ear as a mechanical spectrum analyzer.

To assess human sensitivity to the changes of the HIT slope we used "Change Tempo without Changing Pitch" option of the software, setting the length of the recording in the "Change Tempo" panel to a series of values and replaying the recoding using the "Preview" option.

For the recording of the word "sound" in Fig.1, the changes of the recording length from 0.8s to 0.7s and to 0.9s were clearly recognizable. This means that if the HIT slope does fluctuate over time it does not deviate from some median value more than $\pm 10\%$. A time interval of 1s will be perceived by HIT as being somewhere within 0.9-1.1s.

The software allows changing individually the tempo of the subintervals of the recording in Fig.1. Changing the tempo of different subintervals to different values it is possible to make the replayed word "sound" unrecognizable, though all the frequencies in its recording will remain the same.

Without an accurate enough HIT, hearing could only provide information about presence of sounds of certain frequencies while recognition of the meaning contained in the time sequence of sounds would have been impaired or impossible.

***Internal integration of acceleration***

It is well known that if a blindfolded person is rotated in a swivel chair 5 times over 10 seconds and then stopped the person will erroneously feel that he or she is still rotating, which is a form of vertigo. In this experiment the sensors that provide the motion information are human accelerometers, presumably located in the inner ears. The fact that the person experiences having an angular velocity, which is an integral of the acceleration, suggests that somewhere inside the human body the accelerations are integrated.

The angular velocity and the angular and linear acceleration of the two inner ears in the experiment are schematically shown in Fig.2. Angular velocity, $\omega$, and angular acceleration, $\alpha$, are the same for both inner ears. Tangential, $a$ , and centripetal, $a_c$ , accelerations of the inner ear will occur if there is a non-zero distance, $r$, between the inner ear and the axis of rotation. In general, $a_t$ and $a_c$ are different for the two ears. If a sitting person is rotated around a vertical axis then at least one of the ears will not be on the axis and will have both of these accelerations.

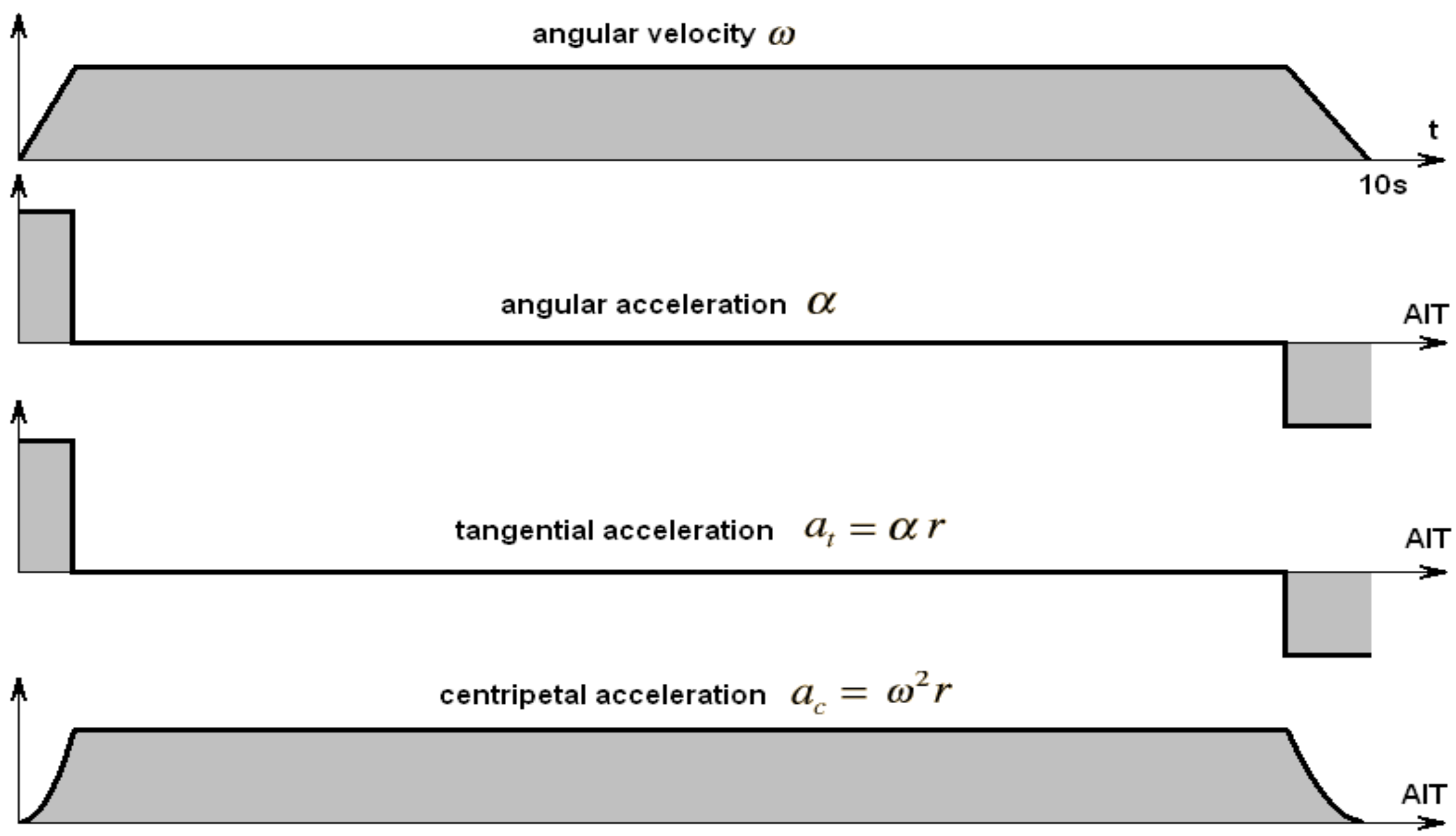

Fig.2. Schematic graphs of the angular velocity and accelerations of the inner ear.

The accelerations shown in Fig. 2 are inputs to human accelerometers. Inside the body these accelerations are integrated into the expected angular velocity, in the case of the spinning chair experiment – erroneously.

On the other hand, if a person turns his or her head 1/8$^{th}$ of the full circle in 1/4$^{th}$ of a second, which would result in the head velocity as in the spinning chair experiment, normally there will be no vertigo. We know this because we habitually turn our heads much faster than this. This means that under some circumstances the angular integration is accurate.

It seems there are two ways in which the change in velocity, $\Delta v = \int a(t)\, dt$, can be calculated inside the body. First way is to register the accelerations as functions of the inner time and have a mechanism of integration of these functions over the inner time. Let us call the inner time that would be needed for this the acceleration inner time, AIT. For this way to work AIT must be a linear function of time, $AIT = kt$. Then $\Delta v = \frac{1}{k}\int a(AIT)\, d(AIT)$, which means that integration over the inner time allows obtaining the required result.

The other way can be rooted in the well-known fact that upon registration by the inner ear the mechanical input, acceleration $a$, is converted into the electrical current, $i = f(a)$ [3, 4]. Suppose this dependence is linear, $i = ka$. The electrical current is the first derivative over time of the electrical charge, $q$, that is passing through the electrical circuit, $i = \frac{dq}{dt}$. Then $\Delta v = \frac{1}{k}\Delta q$. Hence if the acceleration registration mechanism can count the electrical charge that is passing through the circuit then an inner time is not needed to calculate the velocity change because the increment of the velocity is proportional to the increment of the charge.

Conversely, if there is a mechanism that can count the passing electrical charge, e.g. count the ions that are passing through an electrolyte or through the cochlea hair cells, then in the presence of a constant current the count itself can serve as the inner time. The silent stria vascularis current appears to be constant and significant [5]. If the mechanism of the silent current charge count (or its portion passing through the hair cells) exists then the count can serve as the inner time.

Note that the angular and tangential accelerations are about the same, in value and duration, both in the swivel chair experiment and during the short head turn. In this case the only difference between the long and short rotations is in the duration of the centripetal acceleration that is not included into the integration for obtaining the angular velocity. It may be that a prolonged application of the centripetal acceleration drains the resources of the system, resulting in the angular and tangential accelerations registered at the end of the rotation to be imprecise. It would be very interesting to perform an experiment in which the rotational axis goes exactly through both inner ears (e.g. while a person is laying on his side) to study the vertigo effects in the case when only the angular acceleration is present.

### *Location of the hearing inner clock*

As was discussed above, without an inner time the hearing sensory mechanism could only provide information about the presence of a sound and its frequency, but the recognition of complex sounds unfolding in time would have been impossible. This indicates that the hearing sensory mechanism – as complex as it is, – evolutionarily could have been developed only in the presence of an accurate enough inner clock.

On the other hand, are there other processes in an animal that could have prompted the evolutionary development of an accurate enough for hearing inner clock *before* hearing was developed? The answer to this seems to be unknown.

In the absence of a better guess, we speculate that the inner clock needed for hearing was evolutionarily developed together with hearing as a part of the same system. Taken into account that in vertebrates the hearing and acceleration sensory mechanisms are bundled together in the inner ear, and that integration of the accelerations either requires an inner clock or provides a mechanism that can serve as an inner clock, we suggest a hypothesis that the hearing inner clock is located, at least partially, within the same bundle. The acceleration inner clock, if it exists, is also located there.

### *Continuous inner time and sleep*

Let us use IT to denote HIT or AIT (if the latter exists). Suppose that IT(t) is a continuous function of time. For the convenience of the discussion, we assume that the IT slope > 0 so that IT(t) increases over time. Then either the IT(t) range is so large that it is not exceeded during the animal's life span or there must be periods when the function IT(t) decreases.

Let us explore the second possibility (Fig.3).

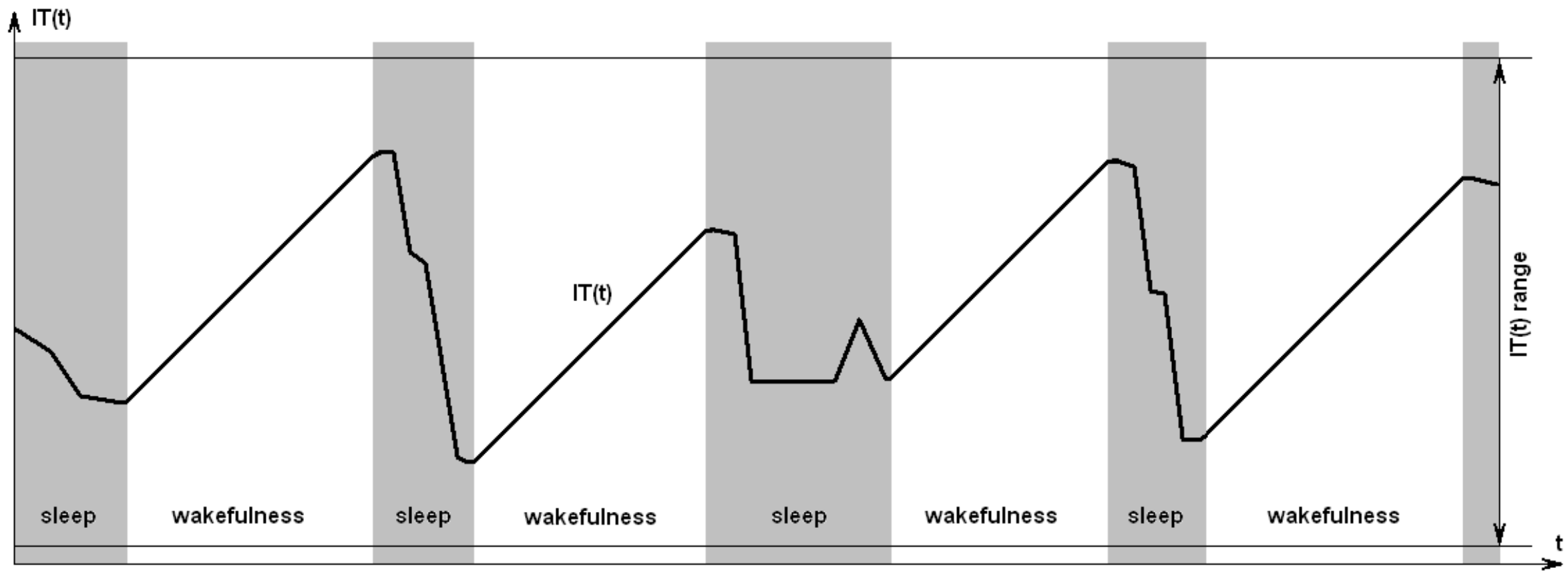

Fig.3. Schematic graph of the continuous inner time, IT(t).

During the periods of time when the fixed IT slope is not maintained and IT(t) decreases the sound recognition is impossible and maintaining balance is difficult because of the misinterpretation of the acceleration inputs. This would be the principle reason why an animal needs to sleep, despite all of the vulnerabilities an animal in the wild experiences when sleeping.

We don't have a theory of how exactly IT(t) would change during sleep but the existence of different stages of sleep hints that it may change unevenly.

On the other hand, if the IT(t) graph in Fig.3 is in principle correct, then it seems that a human should be able to consciously measure, with at least ~10% precision, the time intervals between the events within the same wakefulness period. This does not occur. So either IT(t) is not continuous or it cannot be consciously queried. The latter is very likely as it seems there is no an evolutionary advantage that the exact knowledge of time could provide.

Unrelated to the specific discussion of HIT and AIT, there might be inner clocks/times that serve purposes where the exact proportionality to the physical time does not matter. If there is a physical process or characteristics that serves as an inner time then the inner time will be continuous with respect to the physical time (assuming that all these internal processes or characteristics are continuous). The range of the inner time will be limited and it cannot increase everywhere. When it decreases the inner time will go backwards with respect to the physical time. What was before and after will be scrambled internally. It is difficult to list all the things that could go wrong if the inversed inner time were used, because we don't know all the internal processes where an inner time is used. E.g., creation and usage of the conditioned reflexes would be impossible.

There are many internal processes, such as memory consolidation needed for learning, that occur during sleep [6]. However it is not known why these processes actually require sleep, which is a significant suppression of the animal's sensory inputs. Drawing a comparison with the processing data by a computer, the reorganization of the computer data does not require turning off the computer inputs. However if an animal has a continuous inner time then sleep may be a "mathematical necessity", needed to avoid misinterpretation of the inputs during the inner time "rewinding".

### *“Saw-teeth” hearing inner time*

As discussed above, HIT(t) maybe is not a continuous function of time. Yet it must be a linear function of time, always with the same slope, for at least as long as it is needed to recognize the meaning of a sound. HIT(t) cannot have breaks that would create internal signal sequences that are shorter than it is needed for the recognition. This leads to the following model, Fig.4.

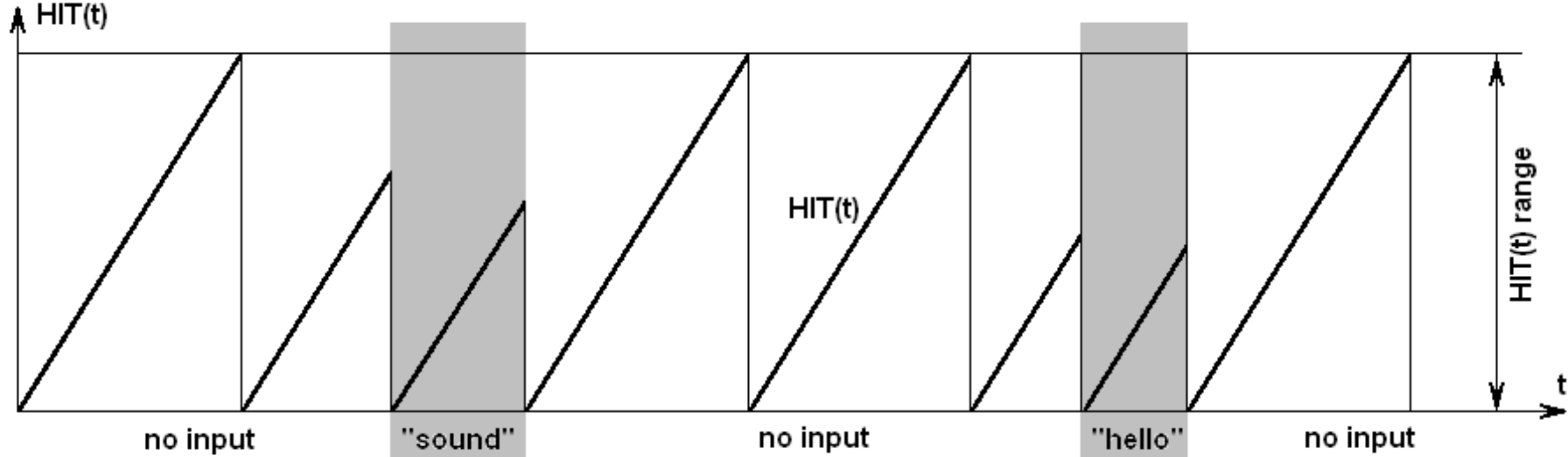

Fig.4. Schematic graph of the non-continuous hearing inner time, HIT.

In the absence of the input sound HIT(t) changes linearly until it reaches its range, then it breaks and resets to 0. Resetting to 0 is also triggered by arrival of new sounds, the words “sound” and “hello” in Fig.4, and by the completion of the sounds recognition.

Within this model the object being recognized by hearing is a function of two arguments with a fixed domain. The first argument is the length along the cochlea spiral to where the electrical signal, which is the value of the function, is produced. First argument’s domain is the whole cochlea spiral length. The second argument is HIT, which is limited to the HIT range.

Humans and some other animals are able to recognize still visual images. In this case the object of recognition is also a function (or several functions, if we include color) of two arguments with a fixed domain, where the arguments are parametric coordinates (in some parametric representation of the retina surface) of the point on the retina from where the signal is coming.

It is possible that similar mechanisms are used for the recognition of the hearing and visual inputs.

### *Inner time and acceleration*

In 1987 the author of this paper was subjected to a strong deceleration during a car crash. The exact parameters of the deceleration are unknown but they must be comparable to the deceleration of a buckled up person in a car that hits a concrete wall at a speed of 70-90 km/hour.

Two most remarkable observations during the crash were, a set of fractures propagating through the windshield so *slowly* that the author could see the details of the fractures development, and a heavy 4.5 liter metal thermos filled with tea that was *slowly* floating in the air near author’s head as if it was a toy balloon.

In reality, both events, the fracture propagation and the thermos’ flight, lasted a fraction of a second. In the author’s inner time they lasted several seconds. This means that under those circumstances the IT slope had temporarily increased, 10-100 times.

There are many evidences of similar experiences, in particular, by the test pilots [7]. There is a research that tries to prove that these experiences are not a result of perception, but a result recollection of a dramatic event [8]. In Appendix 1 we have shown that this research is based on an erroneous interpretation of the experimental data and cannot be relied upon.

On the other hand it is well known that prolonged cycles of small accelerations induce sleep. This is commonly used to rock babies to sleep. Those who travelled long distances in a train sleeping car know that it is applicable to adults as well. As it is noticed in [9], “the nature of the link between rocking and sleep is poorly understood”. If the suggested theory about the bundling of the inner clock with the acceleration sensory mechanism is correct then a possible explanation of the rocking effect may be that a prolonged acceleration load exhausts the whole mechanism, including the inner clock, which would result in sleep.

In both cases, hearing inner time and acceleration inner time, there must exist a mechanism that allows measurement of the short time intervals.

***Conclusions***

We have demonstrated that accurate inner clocks are needed for the recognition of complex sounds. The arguments were made that the clocks or their parts may be located in the inner ear. Experimental verification of the proposed theories is required. Chronic observations that would cover normal wakefulness-sleep cycle may be needed for the inner clock location.

We thank Dr. Alec N. Salt for assistance in research on the cochlea electrodynamics, and Dr. Marat M. Rvachev and Dr. Timur M. Rvachov for critique and support.

Ontario, Canada, 2012

***References***

1. Marc Wittmann, The inner experience of time. Phil. Trans. R. Soc. B 2009 364, 1955- 1967,doi: 10.1098/rstb.2009.0003, http://rstb.royalsocietypublishing.org/content/364/1525/1955.full.pdf
2. http://audacity.sourceforge.net/
3. Alec N. Salt, Cochlear Fluids Research Laboratory, http://oto2.wustl.edu/cochlea/
4. Philine Wangemann, Supporting sensory transduction: cochlear fluid homeostasis and the endocochlear potential. J Physiol 576.1 (2006) pp 11–21, http://jp.physoc.org/content/576/1/11.full.pdf
5. M Zidanic, W E Brownell, Fine structure of the intracochlear potential field. I. The silent current. Biophys J. 1990 June; 57(6): 1253–1268. http://www.ncbi.nlm.nih.gov/pmc/articles/PMC1280835/
6. Susanne Diekelmann and Jan Born, The memory function of sleep. Nature Reviews Neuroscience 11, 114-126 (February 2010) | doi:10.1038/nrn2762 http://www.nature.com/nrn/journal/v11/n2/full/nrn2762.html
7. https://www.e-reading.club/chapter.php/1000632/24/Shelest_Igor_-_Lechu_za_mechtoy.html#n_3
8. Chess Stetson, Matthew P. Fiesta, David M. Eagleman, Does Time Really Slow Down during a Frightening Event? https://journals.plos.org/plosone/article?id=10.1371/journal.pone.0001295
9. Laurence Bayer, Irina Constantinescu, Stephen Perrig, Julie Vienne, Pierre-Paul Vidal, Michel Mühlethaler and Sophie Schwartz, Rocking synchronizes brain waves during a short nap. Current Biology, Volume 21, Issue 12, R461-R462, 21 June 2011, doi:10.1016/j.cub.2011.05.012, http://www.cell.com/current-biology/abstract/S0960-9822(11)00539-2